\documentclass{jfm}
 \pdfoutput=1 
\usepackage{graphicx}
\usepackage{epstopdf, epsfig}
\usepackage{amsmath,makecell,hyperref,cleveref,url}
\usepackage{xcolor}
\hypersetup{
    colorlinks,
    linkcolor={red},
    citecolor={blue},
    urlcolor={blue}
}

\definecolor{RoyalBlue}{RGB}{0,0,128}

\shorttitle{Spectral analysis of mixing in 2D flows}
\shortauthor{H. Arbabi and I. Mezi\'{c}}

\graphicspath{{./}{./figures/}}

\title{Spectral analysis of mixing in  2D high-Reynolds flows}

\author{Hassan Arbabi\aff{1,2}
  \corresp{\email{arbabiha@gmail.com, mezic@ucsb.edu}},
Igor Mezi\'c\aff{2}}

\affiliation{\aff{1}Department of Mechanical Engineering, Massachusetts Institute of Technology
\\Cambridge, MA 02139, USA
\aff{2}Department of Mechanical Engineering, University of
California Santa Barbara \\Santa Barbara, CA 93106, USA }

\begin{document}

\maketitle

\begin{abstract}
{	
 We use spectral analysis of Eulerian and Lagrangian dynamics to study the advective mixing in an incompressible 2D bounded cavity flow. A significant property of such a rotational flow at high Reynolds numbers is that mixing in its core is slower than wall-adjacent areas and corner eddies. 
We explain this property by considering the resonance between frequencies of unsteady motion --- captured by the Koopman spectral analysis of the velocity field --- and the circulation frequency of Lagrangian tracers in the mean flow. In high-Reynolds rotational 2D flows, the vorticity in the rotational core is uniformly distributed, which leads to uniform distribution of circulation periods in the mean flow, i.e., the kinematics in the core of mean flow is like rigid-body rotation. 
When this ``rigid" core is exposed to velocity fluctuations arising from bifurcations
at high Reynolds, it shows more resilience toward resonance in Lagrangian motion and
hence mixes more slowly compared to other areas.
We also show how our qualitative resonance argument extends to chaotic flows where the classical tools of dynamical systems are not applicable.
}


\end{abstract}
\begin{keywords}
chaotic advection, Prandtl-Batchelor theorem, Koopman mode decomposition, proper orthogonal decomposition
\end{keywords}

\section{Introduction}
Mixing is an important aspect of many natural and industrial flows.
Characterizing the vertical mixing in the ocean and atmosphere, for example, constitutes the main challenge in modeling the earth climate \cite[]{large1994oceanic,sherwood2014spread}, while understanding the horizontal mixing on the ocean surface helps us predict the movement of pollution and could lead to more effective strategies for containment \citep{coulliette2007optimal}.
Other examples from natural flows include the mixing in the earth mantle which led to formation of oceanic islands with nonuniform geochemistry \citep{ferrachat2001mixing},  blood flow mixing in relation to health and disease \citep{shadden2008characterization} and the role of mixing in shaping the ecological equilibrium in oceanic environment \citep{valentine2012dynamic}.
For the industrial flows, on the other hand, we often try to manipulate mixing, e.g., design devices that efficiently mix the fluids given the constraints by the specific application \citep{khakhar1987case,stroock2002chaotic}.
All such efforts are based on our understanding of many factors that play a role in mixing like the flow dynamics, device geometry and initial configuration of the mixing fluids.

Study of mixing in real-world problems is difficult. Most of rigorous analysis in this field comes from the theory of chaotic advection, which treats the motion of tracers in the flow as  a dynamical system \citep{aref1984,ottino1989kinematics,rom1990analytical}. This theory has had a great influence on how we view the transport of material in flows which are steady or time-periodic, but it offers significantly less insight for flows which are \emph{aperiodic} in time. On the other hand, most of the natural flows and many industrial flows show aperiodic time dependence. As a result, a large number of techniques have been devised to fill the gap between the knowledge of mixing in periodic flows and the aperiodic flows that appear in practice. Most of these techniques strive to characterize the mixing in a given aperiodic flow by detecting and visualizing the flow structures which are most informative about the collective behavior of Lagrangian trajectories. Those techniques, just to name a few, include the theory of Lagrangian coherent structures \citep{Haller2015}, topological analysis via braid dynamics \citep{boyland2000topological,budisic2015finite}, theory of finite-time coherent sets \citep{froyland2010transport}, and ergodic analysis by time averaging \citep{poje1999geometry}.

{
In this work, we study the connection between the flow dynamics, i.e., the time evolution of the velocity field, and advective mixing of passive tracers in a 2D flow. In particular, we apply our analysis to lid-driven flow in a square cavity obtained from numerical simulation.
This flow bifurcates into periodic, quasi-periodic and ultimately chaotic dynamics with the increase of Reynolds numbers.
 An important feature of these time-dependent flows is that at very high Reynolds numbers the vorticity approaches a uniform distribution in their rotating core. This is known as Prandtl-Batchelor theorem 
\citep{prandtl1904uber,batchelor1956steady} and was recently extended to unsteady flows \citep{arbabi2019prandtl}. Here, we show that this uniform distribution of vorticity leads to uniform distribution of Lagrangian time periods in the mean flow. This rigid-body type of rotation reduces the chance of resonance with unsteady components of the flow and therefore leads to weaker mixing in the core, compared to areas adjacent to the walls and secondary vortices. 
This behavior, which also holds for quasi-periodic and chaotic flows, is in stark contrast to low-Reynolds flows where mixing is typically stronger away from the walls. 
}

We also investigate how the complexity of mixing process is changed while the temporal regime of the lid-driven cavity flow changes from steady to aperiodic. We use a combination of Koopman Mode Decomposition (KMD) and Proper Orthogonal Decomposition (POD) to extract the hierarchy of the energetic modes in the flow, and use subsets of those modes to build finite-dimensional projected models to approximate the flow evolution. 
Then we quantify how mixing in those models mimic the mixing in the actual flow.  
This will verify the validity of modal analysis in study of mixing and characterize the dimensionality of models required to replicate mixing in complex flows.

\subsection{Contributions and organization of the paper}
Most of the previous work on dynamical-systems analysis of mixing relies on prescribed velocity fields motivated by laminar flows, where the time dependence is usually a result of external forcing. The major contribution of this work is the combination of Navier-Stokes dynamics, in the form of a general constraint on vorticity distribution, and a qualitative resonance analysis from chaotic advection to predict the structure of mixing. Although we closely study the toy example of square cavity flow, our arguments are independent of the specific geometry and only depend on presence of a rotational core far from the walls. Therefore, we expect the general conclusion, i.e. weaker mixing in the core, to hold for all high-Reynolds 2D rotational flow structures. 

We employ a qualitative analysis of resonance to understand the strength of mixing in various regions of the flow, and this analysis makes valid predictions for the aperiodic flow as well. Hence it shows great promise for analysis of real-world flows where many methods of classical dynamical systems theory, which are well-defined only for steady and periodic flows, cannot be helpful.
Throughout our analysis, we also heavily use the concept of breaking down the flow to its oscillating modes. The results on mixing of projected models indeed support this approach, namely, that study of mixing-related phenomena in complex flows can be approximated by using a sufficient number of robust modes that constitute the bulk of the flow energy.

The rest of this paper is organized as follows: we first give an account of the previous studies on mixing in lid-driven cavity flow. In \cref{sec:cavityKMD}, we specify the geometry and dynamics of the lid-driven cavity flow in our study. We summarize our previous work \citep{arbabi2017study} by reviewing the spectral analysis of the flow field for each temporal regime that appears due to the flow bifurcations with the increase of Reynolds number. In \cref{sec:mixing_tools}, we review our tools for mixing analysis:
 \Cref{sec:hypergraphs} reviews the application and computation of hypergraphs for visualization of mixing in a given flow. \Cref{sec:Mix-norm} discusses quantitative analysis of mixing using the mix-norm.
 We present the qualitative picture of mixing in cavity flow in \cref{sec:PBeffect} and discuss the role of Prandtl-Batchelor theorem leading to weak mixing in the core.  In  \cref{sec:approx}, we describe our experiment for analysis of modal contributions to mixing and present its results.

\subsection{Previous studies on mixing in lid-driven cavity flow}
The mixing in the lid-driven cavity flow has been explored from many aspects and under different settings. This flow requires a simple computational setup and it is regularly used  as a computational benchmark problem \citep{ghia1982high}. It is also studied via experiments in 2D \citep{gharib1989liquid}  and 3D geometry \citep{koseff1984}.
The 2D flow which we focus on here represents a simplified model of geophysical flows driven by surface shear \citep{tseng2001mixing}, or common types of mixers in polymer engineering \citep{chella1985fluid}. Most of the previous studies have focused on low-Reynolds cavity flows with time-dependent lid motion, and investigated the effect of different factors like lid motion frequency and cavity geometry on enhancement of mixing. 

In the steady lid-driven cavity flow, which is the stable solution at low Reynolds, mixing is generally poor since the tracers are confined to move along fixed streamlines \citep{ottino1989kinematics}. The experiments by \citet{chien1986laminar} and \citet{leong1989experiments} showed that mixing is greatly improved if periodic lid motion is used to generate periodic flow. In that case, the motion of tracers inside the cavity is comprised of both periodic and chaotic trajectories. The chaotic trajectories make the well-mixed regions while the tracers with periodic motion form coherent patches of fluid called \emph{periodic islands}. These islands prevent full mixing because the fluid blobs remain trapped inside them and do not spread.
The experiments showed that size of these islands are dependent on the forcing frequency, and it was understood that there are optimal frequencies, at which, the islands would vanish and complete mixing would occur.
\citet{ling1992mixing} and \citet{ling1993effect} identified such frequency ranges by studying the linear stability of the periodic orbits corresponding to those islands. Their results showed good agreement with simulation and previous experiments, and motivated further studies on detection of periodic orbits in the cavity and their role in mixing \citep[see e.g.][]{jana1994experimental,meleshko1996periodic,anderson1999analysis,anderson2000chaotic,stremler2007generating}.

The aperiodic mixing is much less explored in the studies regarding the lid-driven cavity flow. \citet{franjione1989symmetries} and \citet{ottino1992chaos} proposed a non-random aperiodic protocol for the lid motion to enhance mixing. The underlying idea in their work is to manipulate the symmetries in the flow to break up the periodic islands.
The numerical studies  by \cite{liu1994quantification} also showed that the aperiodic lid motion can lead to stronger and more uniform mixing in the cavity flow. We note that the aperiodicity of the flow in the above studies is generated by modulating the lid velocity, whereas in our study, the aperiodicity arises due to Navier-Stokes dynamics at high Reynolds number while the lid velocity is constant.

There are also a number of studies that investigated the mixing in lid-driven cavity flow while considering the effect of different geometries \citep{ottino1992chaos,jana1994chaotic,migeon2000effects}, flow stratification \citep{tseng2001mixing}, multi-phase flow configuration \citep{chakravarthy1996mixing,chella1996mixing} and motion of freely moving solid bodies within the flow \citep{vikhansky2003chaotic,hwang2005chaotic,pai2013numerical}.

\section{Dynamics of lid-driven cavity flow}\label{sec:cavityKMD}
We use a data-driven approach based on the spectral analysis of Koopman operator to characterize and present the cavity flow dynamics. 
The idea of Koopman operator goes back to  \cite{Koopman1931}, while its application for data-driven analysis of high-dimensional systems was proposed just in the last decade \citep{mezic2004comparison,Mezic2005}. In particular, \cite{Mezic2005} introduced the notion of Koopman modes, which are analogues of eigenvectors in linear systems, for nonlinear evolution of spatio-temporal systems like fluids.  \citet{rowley2009} pioneered the data-driven Koopman approach for fluid flows and pointed out the connection between Koopman mode decomposition (KMD) and the dynamic mode decomposition algorithm \citep{Schmid2010}. In a previous work, we used KMD to categorize and study different dynamic regimes of the cavity flow \citep{arbabi2017study}. The key to understanding the flow regime (i.e the geometry of attractor in the state space of the flow)  is the Koopman spectrum, while the Koopman modes reveal the associated spatial structures in the flow domain.
This section summarizes the results of that work  by recounting the  sequence of flow regimes that appear with the increase of Reynolds number in the cavity flow.

The cavity flow domain is a square box $[-1,1]^2$ with fixed walls except at the top where the wall moves with the velocity profile
\begin{equation}
u_{top}(x)= (1-x^2)^2,\quad -1\leq x\leq 1.
\end{equation}
We define the Reynolds number as
\begin{equation}
\Rey= \frac{L_cU_c}{\nu}=\frac{2}{\nu},
\end{equation}
where $L_c=2$ and $U_c=1$ are the characteristic length and velocity respectively, and $\nu$ denotes the fluid kinematic viscosity in the numerical simulation. 

\begin{figure}
	\centerline{\includegraphics[width=1 \textwidth]{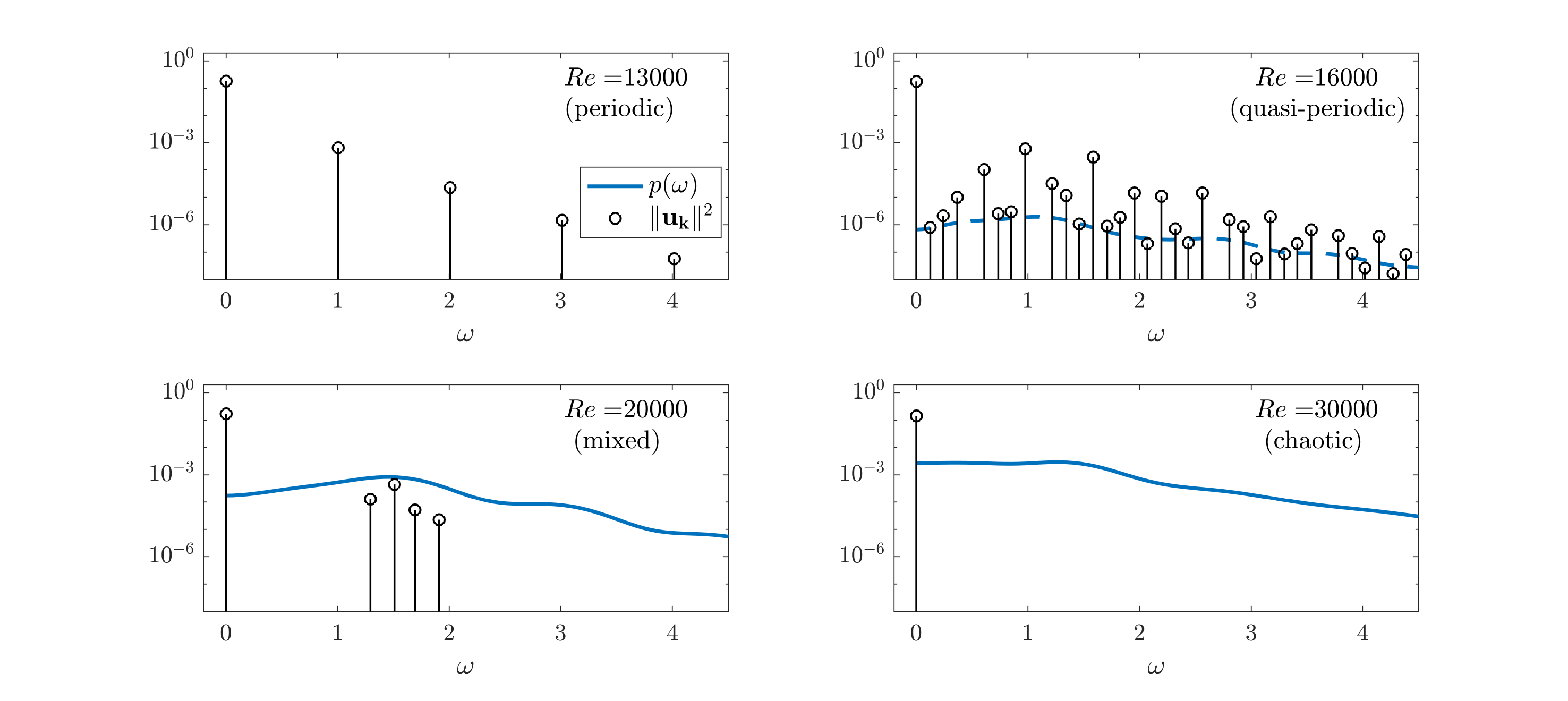}}
	\caption{Flow dynamics revealed by Koopman spectrum of data: In periodic and quasi-periodic flow the Koopman frequencies (circles) are multiples of the one or two basic frequencies (top panels), in chaotic flow the only Koopman discrete frequency is zero and the rest of energy is in the continuous spectrum (bottom right), and for cavity flow there exists an intermediate range with mixed spectra (bottom left). Some of the panels are adapted from \citet{arbabi2017study}.}
	\label{fig:spectrum}
\end{figure}

\begin{figure}
	\centerline{\includegraphics[width=1\textwidth]{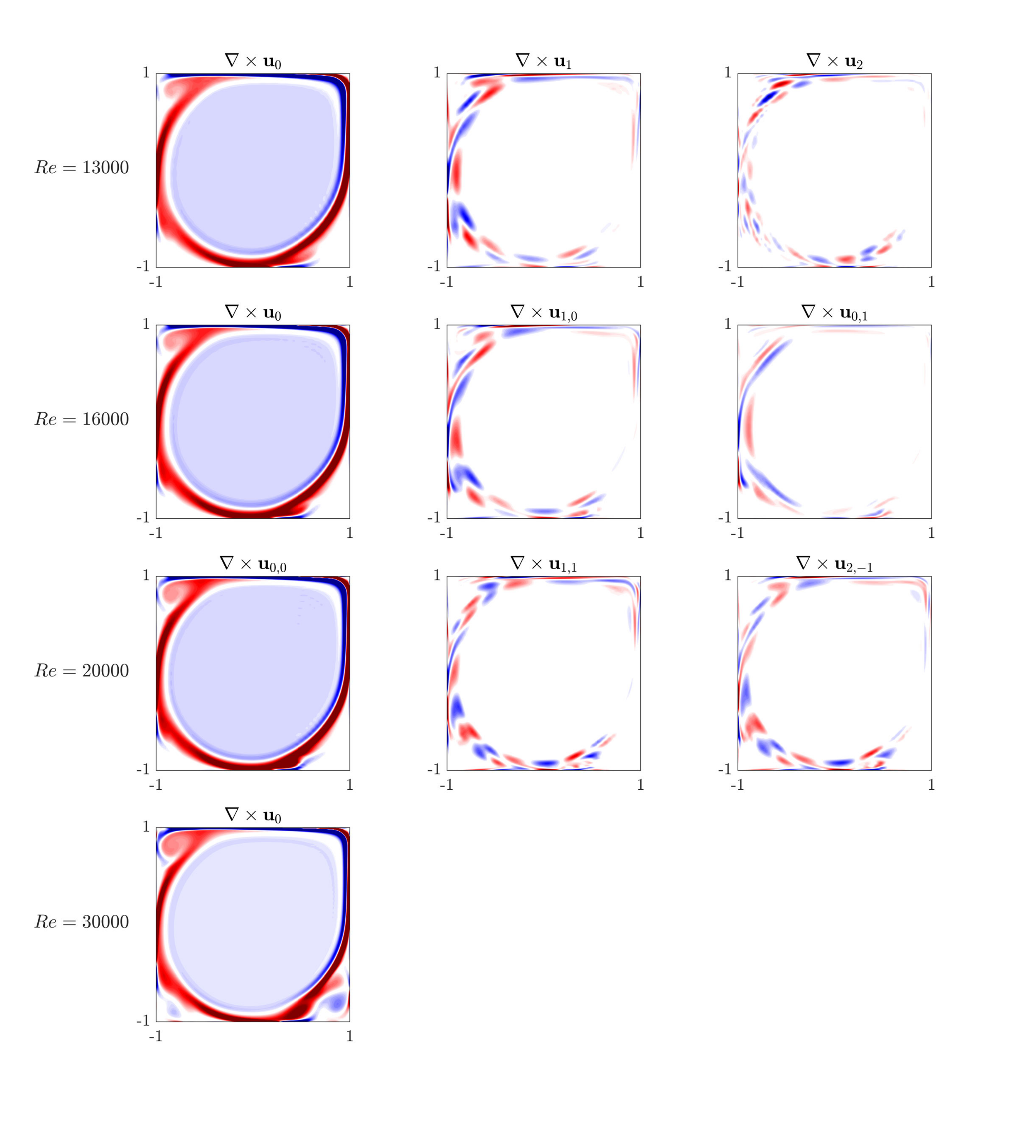}}
	\caption{The (real part of) Koopman modes of vorticity for cavity flow. Clockwise rotation is marked with blue and counterclockwise with red. Some of the panels are adapted from \citet{arbabi2017study}.}
	\label{fig:Kmodes}
\end{figure}

\Cref{fig:spectrum} summarizes the dynamics of the cavity flow as explained by the Koopman spectrum of flow field data.
For low Reynolds numbers, the flow started from zero initial condition converges to a steady solution. At a Reynolds number slightly above $10000$, a Hopf bifurcation occurs and the asymptotic flow dynamics becomes periodic, i.e., the trajectory in the state space of the flow converges to a limit cycle. The Koopman frequencies for the periodic flow consist of a basic frequency $\omega_1$ (frequency of limit cycle in the state space) and its multiples $\omega_k:= k\omega_1$ with $k\in\mathbb{Z}$. The evolution of velocity field with time is described by the KMD of the form
\begin{equation}
\mathbf{u}^p(x,y,t)= \mathbf{u}_0(x,y) + \sum_{\substack{k\in \mathbb{Z}\\k\neq 0}} \mathbf{u}^p_k(x,y) e^{ik\omega_1 t}, \label{eq:KMDperiodic}
\end{equation}
where $\mathbf{u}_k$ is the Koopman mode of velocity field associated with Koopman frequency $k\omega_1$ and $\mathbf{u}_0$ is the mean flow (i.e. the Koopman mode associated with zero frequency). Three Koopman modes of the periodic flow with highest kinetic energy are depicted in the first row of \cref{fig:Kmodes}.

At a Reynolds number slightly above $15000$, the flow undergoes another Hopf bifurcation and becomes quasi-periodic. The quasi-periodic flow exhibits two basic frequencies $\boldsymbol\omega= (\omega_1,\omega_2)$, and the Koopman frequencies are the linear combination of the basic frequencies with integer coefficients $\mathbf{k}=(k_1,k_2)$, i.e., $\omega_{\mathbf{k}}:=k_1\omega_1+k_2\omega_2$ with $k_1,k_2\in\mathbb{Z}$. The KMD for the quasi-periodic velocity field is given by
\begin{equation}
\mathbf{u}^{q}(x,y,t)= \mathbf{u}_0(x,y) + \sum_{\substack{\mathbf{k}\in \mathbb{Z}^2\\ \mathbf{k}\neq(0,0)}} \mathbf{u}_\mathbf{k}(x,y) e^{\mathbf{k}\cdot\boldsymbol{\omega} t},
\end{equation}
with the notation similar to \eqref{eq:KMDperiodic}. \Cref{fig:Kmodes} shows the mean flow and the basic oscillatory Koopman modes (i.e modes associated with the two basic frequencies). 

As the Reynolds number is increased higher than 18000, the Koopman spectrum becomes mixed, that is, a combination of discrete frequencies that describe the quasi-periodicity and the continuous spectrum which corresponds to the chaotic components of the velocity field. The Koopman mode decomposition for this flow reads
\begin{equation}
\mathbf{u}^{m}(x,y,t)= \mathbf{u}_0(x,y) +  \sum_{\substack{\mathbf{k}\in \mathbb{Z}^2\\ \mathbf{k}\neq(0,0)}} \mathbf{u}_\mathbf{k}(x,y) e^{\mathbf{k}\cdot\boldsymbol{\omega} t} + \mathbf{u}_c(x,y,t), \label{eq:KMDmixed}
\end{equation}
with $\mathbf{u}_c$ denoting the chaotic component of the velocity field. 
Note that for this flow it is difficult to detect the true modes within the background chaos and \cref{fig:spectrum}  shows the modes that have passed the robustness test described in \citet{arbabi2017study}.
In the same figure, the strength of the continuous spectrum and chaotic component is reflected by the Koopman spectral density $p(\omega)$  which is similar to power spectral density of stationary stochastic processes. 
For more on continuous spectrum expansion see  \citet{Mezic2013analysis}.

As the Reynolds number is increased even further, the continuous spectrum becomes more dominant so that no quasi-periodic components can be found for Reynolds numbers above 22000. Then the KMD consists only of the mean flow and the chaotic component, i.e.,
\begin{equation}
\mathbf{u}^{a}(x,y,t)= \mathbf{u}_0(x,y) +   \mathbf{u}_c(x,y,t). \label{eq:KMDaperiodic}
\end{equation}

There are two observations of Koopman modes of the flow which are important for study of mixing.
First, vorticity in the core of the mean flow is constant and it is zero in the core of oscillatory modes (\cref{fig:Kmodes} and \ref{fig:Pmodes}). This is a consequence of Prandtl-Batchelor theory and we will discuss it in the results. The second observation is that
 linear relationships between different observable fields carries over to the Koopman modes of those observables. For example, the stream function $\psi$ and velocity field $\mathbf{u}$ are related via the linear operator $\nabla^\perp:=[\partial/\partial y,-\partial/\partial x]^T$ such that $\mathbf{u}= \nabla^\perp \psi$. Now if we let $\psi_j$ and $\mathbf{u}_j$ denote the Koopman modes of these two observable fields associated with Koopman frequency $\omega_j$, then
\begin{equation}
\mathbf{u}_j= \nabla^\perp \psi_j.
\end{equation}
This also implies that the KMD of those observables contain the same Koopman frequencies (as long as no $\psi$-mode is in the null space of $\nabla^\perp$), and the Koopman modes describe the same steady flow field. Therefore, when we study the effect of the Koopman mode associated with frequency $\omega_j$ on mixing, we can switch
between different representations of that mode including the stream function mode, velocity mode and also vorticity mode. Note that in this work, we identify complex conjugate pairs of Koopman modes with their common frequency.

\subsection{POD of chaotic components}

\begin{figure}
	\centerline{\includegraphics[width=1 \textwidth]{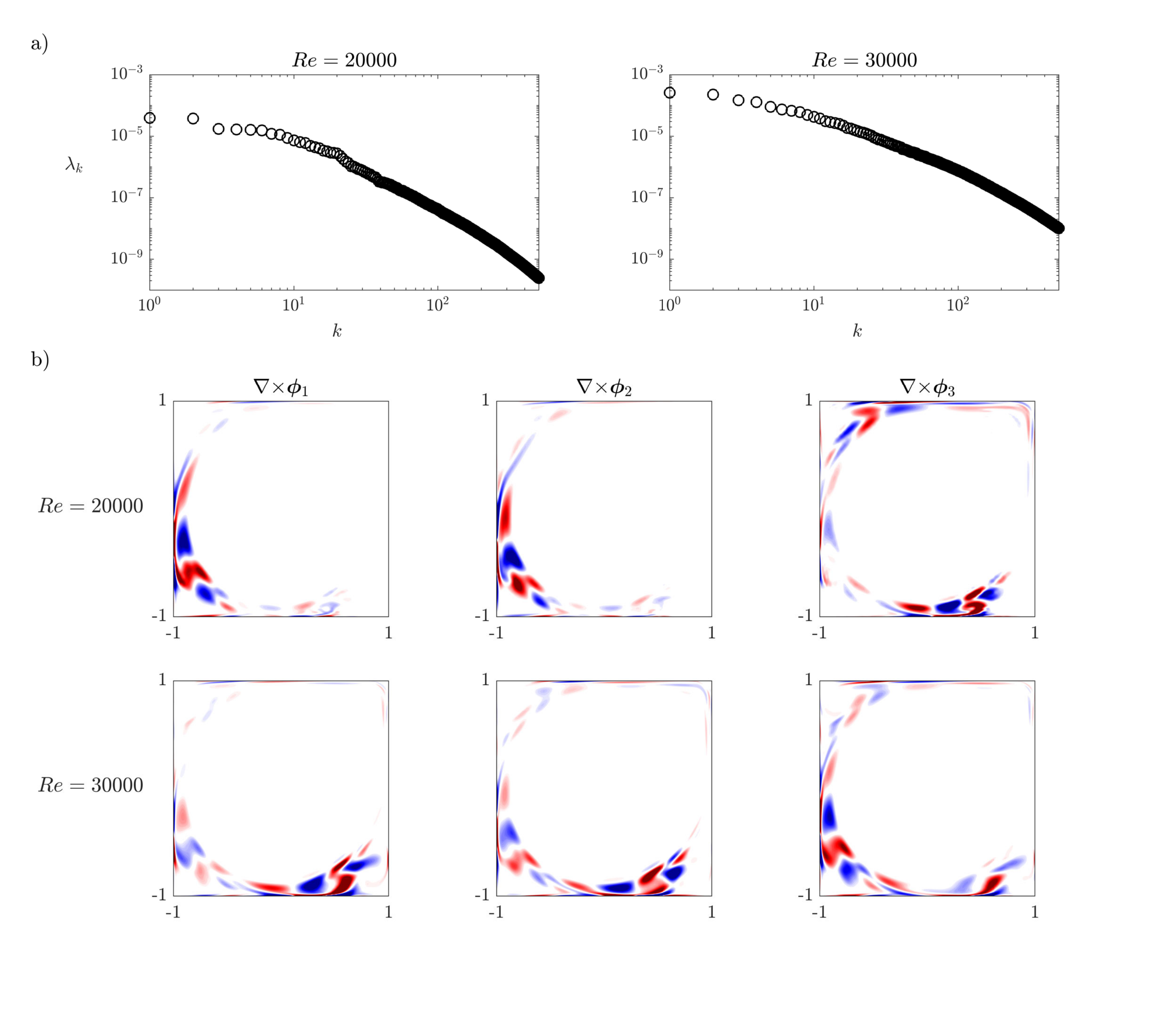}}
	\caption{POD of the chaotic component: a) POD eigenvalues (i.e. kinetic energy of normalized POD modes), b) the vorticity field of 3 most energetic POD modes.
	The number of POD modes (in addition to Koopman modes of quasi-periodic part) required to resolve \%99 of unsteady kinetic energy is 43 and 186 for $Re=20000$ and $Re=30000$ respectively.
	}
	\label{fig:Pmodes}
\end{figure}

The chaotic part of the velocity field, which is present in the high-Reynolds flows, does not admit a decomposition into Koopman modes. However, the chaotic component of a post-transient flow (i.e. evolving on the attractor) is a realization of a stationary stochastic process and therefore we can use statistical tools such as POD to obtain a meaningful decomposition.
POD is a linear decomposition of the flow field into spatially orthogonal modes and uncorrelated temporal coefficients \citep{holmes2012turbulence}. The POD for the chaotic component $\mathbf{u}_c(x,y,t)$ is denoted by
\begin{equation}\label{eq:POD}
  \mathbf{u}_c(x,y,t)= \sum_k a_k(t) \boldsymbol{\phi}_k(x,y),
\end{equation}
where $ \boldsymbol{\phi}_k$'s are the POD modes and $a_k$'s are the time-dependent POD coordinates. The POD spectra and modes for chaotic components of cavity flow are shown in \cref{fig:Pmodes}.

 Similar to KMD, POD is a linear expansion, but there are two major differences: POD coordinates, unlike Koopman mode coordinates, do not necessarily evolve exponentially with time. Also, POD modes are spatially orthogonal while the Koopman modes are not necessarily so. POD provides a robust representation of post-transient chaotic components 
since the modes become independent of the initial condition due to ergodicity. We will use this representation along with the KMD of quasi-periodic component  to approximate the effect of modes on advective mixing in the flow.

\section{Hypergraphs and mix-norm}\label{sec:mixing_tools}
In this work, we study mixing as the pure advection of passive tracers and scalar fields with the flow. In other words, we focus on the mixing process at the limit of infinitely large \emph{Peclet} number and zero \emph{Stokes} number. To this end, we use two different tools: We employ \emph{hypergraphs} to visualize mixing of each flow and do qualitative analysis. We use \emph{mix-norm} to quantify the contribution of Koopman and POD modes  to mixing of the cavity flow.

\subsection{Hypergraphs}\label{sec:hypergraphs}

The hypergraphs are visualizations of a scalar field known as \emph{mesohyperbolicity} which partitions the flow domain according to the type of Lagrangian deformation  \citep{mezic2010new}. The field of mesohyperbolicity is defined as follows:
\noindent Consider the trajectory of a passive tracer passing through $\mathbf{x}$ at time $t_0$. We denote by $\mathbf{u^*}_{t_0}^{t_0+T}(\mathbf{x})$ the time-averaged Lagrangian velocity of the tracer over the time interval $[t_0,t_0+T]$.
The mesohyperbolicity field, $M$, is defined as the gradient of this averaged velocity with respect to the initial location, i.e., 
\begin{equation}
M(\mathbf{x},t_0,T)\equiv\det\left| \nabla_\mathbf{x} \mathbf{u^*}_{t_0}^{t_0+T}(\mathbf{x})\right|. \label{eq:mesohyper}
\end{equation}
This field uniquely determines the type of Lagrangian fluid deformation in the neighborhood of the tracer. In particular,

1. In the regions where $M<0$, the small patch of fluid around the tracer will be stretched in one direction and contracted in the other while moving during the next $T$ seconds. This deformation is similar to behavior of trajectories in vicinity of a hyperbolic fixed point in a plane, hence called \emph{mesohyperbolic}.

2. When $0\leq M\leq4/T^2$, the fluid patch undergoes rotation while traveling. We call this behavior \emph{mesoelliptic}.

3. The regions with $M>4/T^2$ show the combination of the above deformations, i.e., the fluid patch rotates while it is stretched in one direction and contracted in the other. This type of deformation is called \emph{mesohelical}.

\begin{figure}
  \centerline{\includegraphics[width=1.1 \textwidth]{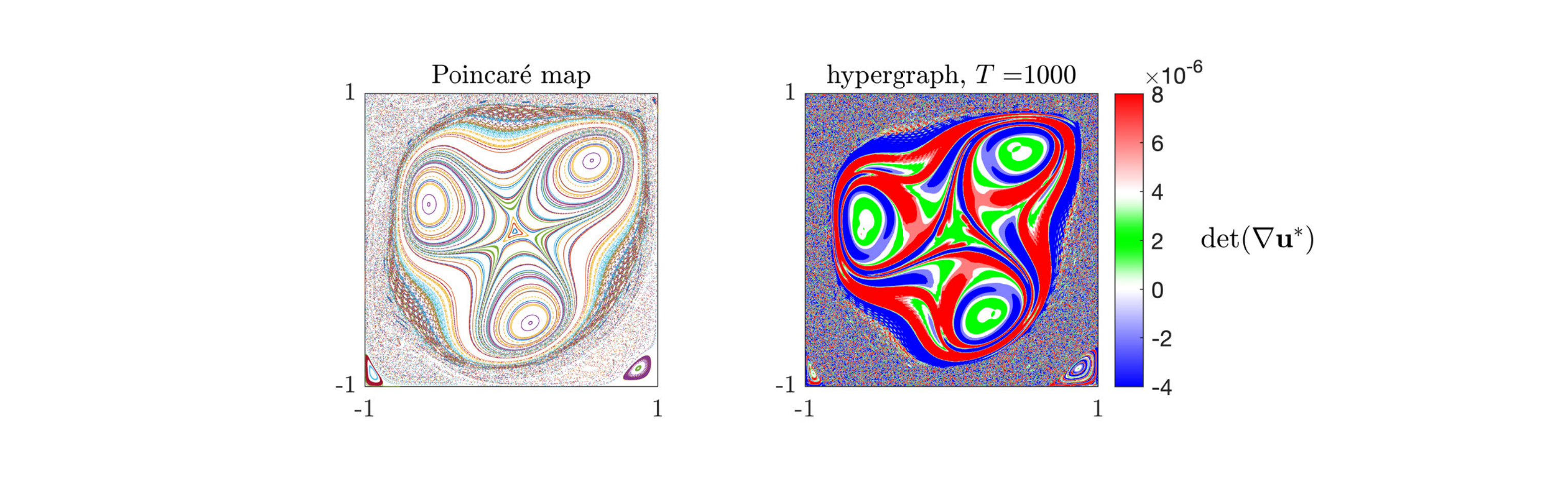}}
  \caption{(a) Poincar\'e map of periodic cavity flow at $\Rey=13000$, computed using 300 random trajectories over 2000 time periods. (b) Hypergraph of the same flow over 1000 sec, computed on a grid of $300\times300$ initial conditions. Blue, red and green colors correspond to mesohyperbolic, mesohelical and mesoelliptic behavior respectively. The (chaotic) mixing zones stand out in the hypergraph as regions with fine mixtures of red and blue.}
\label{fig:standardmap} 
\end{figure}

In the hypergraphs plotted in this paper, the mesohyperbolic behavior is marked by blue, mesoelliptic by green, and mesohelical by red. A  hypergraph of the periodic cavity flow is shown in fig. \ref{fig:standardmap} (right).
A comparison with the Poincar\'e map of the same flow (left in the same figure) shows how hypergraphs can be used to qualitatively assess mixing: The well-mixed regions are revealed in hypergraphs as areas with a fine-grained mixture of the mesohyperbolic and mesohelical deformation (red and blue) ---  similar to hyperbolic sets in dynamical systems theory.
On the other hand, the islands of periodic motion which correspond to poorly mixed regions stand out as concentric bands of alternating colors. 

\citet{mezic2010new} introduced the hypergraphs to study finite-time (or aperiodic) mixing in 2D incompressible flows which cannot be studied via Poincare maps.
Over finite time intervals, the regions of substantial mixing stand out in hypergraphs as areas with a fine-grained combination of red and blue because those regions host an extensive amount of stretching and folding of the material elements which resembles the classic notion of chaotic motion.
On the other hand, the poorly mixed regions divide into two subgroups: regions that are consistently mesoelliptic and therefore show rotation and stagnation zones, and regions with a dominant type of either mesohyperbolic (blue) or mesohelical (red) which denote likely passages for tracer motion in the form of coherent blobs.  \cite{Budisic2016c}  provides a more detailed discussion of hypergraph analysis and its extension to 3D geometry. Here we use hypergraphs to distinguish regions of chaotic mixing from regions of slow mixing such as periodic islands.
For details on computation and visualization of hypergraphs see the \emph{Appendix}. 

\subsection{mix-norm}\label{sec:Mix-norm}
We will use the evolution of a generic initial scalar field to see how well the modal approximations of the flow approximate the mixing process.
Consider the scalar field $c(\mathbf{x},t)$ which could, for example, represent the concentration of a dye. In absence of diffusion, $c$ evolves in time due to the advection by the velocity field, i.e.,
\begin{equation}\label{eq:advection_field}
  \frac{\p c}{\p t} + \mathbf{u}\cdot \nabla c=0.
\end{equation}
We compute the time evolution of $c$ using a semi-Lagrangian scheme: To compute the advection of $c$ over the time interval $[t_1,t_2]$, we  put a regular grid on the flow domain at time $t_2$, and advect the grid points backward in time to  $t_1$. Then we interpolate the field of $c(\mathbf{x},t_1)$ onto the advected grid points. We trace back each advected point to its location on the original grid at time $t_2$ and assign the interpolated value to that location. The semi-Lagrangian scheme harbors much less numerical diffusion than PDE discretization methods and therefore  it enables a robust approximation of norms that are used for characterization of advective mixing. 

There are a few measures to quantify the \emph{mixedness} of a concentration field. In particular, the class of Sobolev-space norms with negative indices are the most popular choice for study of advection-dominated mixing.  Here, we use the earliest version of  such a norm introduced in \cite{mathew2005multiscale}.
 Let
\begin{equation}\label{eq:CFourier}
  c(\mathbf{x}) = \sum_{\mathbf{k}\in\mathbb{Z}^2}c_\mathbf{k} f_\mathbf{k}(\mathbf{x})
\end{equation}
be the standard Fourier expansion of $c$ over the (bounded) flow domain. For the lid-driven cavity, $\mathbf{k}$ is the 2-vector of wave numbers, $c_\mathbf{k}$'s are the Fourier coefficients, and the Fourier functions are
\begin{equation}\label{eq:FourierFuns}
   f_\mathbf{k}(\mathbf{x}) = e^{i\pi( k_1x + k_2y)}.
\end{equation}
Then the mix-norm of $c$ is defined a
\begin{equation}\label{eq:MixNorm}
   \Phi(c)=\bigg(\sum_{\mathbf{k}\in\mathbb{Z}^2}\frac{1}{\sqrt{1+\pi\|\mathbf{k}\|^2}}c^2_\mathbf{k}\bigg)^{1/2},
\end{equation}
which is similar to $l^2$-norm of Fourier coefficients except that coefficients associated with higher wave numbers have smaller weights.
The essential feature of this norm is that it treats mixing as a multi-scale phenomena and puts less weight on smaller spatial scales (i.e. the weights decay as $\mathbf{k}$ increases). When the scalar $c$ is being mixed by the flow, its mix-norm decreases. An intuitive reasoning for this is that the process of mixing stretches and folds the fluid blobs represented by large wave numbers into elongated and tightly spaced filaments represented by small wave numbers. We will use the above mix-norm to characterize the difference of mixing in the field $c(\mathbf{x},t)$  according to different modal approximations of the same flow.
A more detailed review of mix-norms is offered by \citet{thiffeault2012using}.

\section{Slow mixing in the core and Prandtl-Batchelor theorem} \label{sec:PBeffect}
\begin{figure}
	\centerline{\includegraphics[width=1 \textwidth]{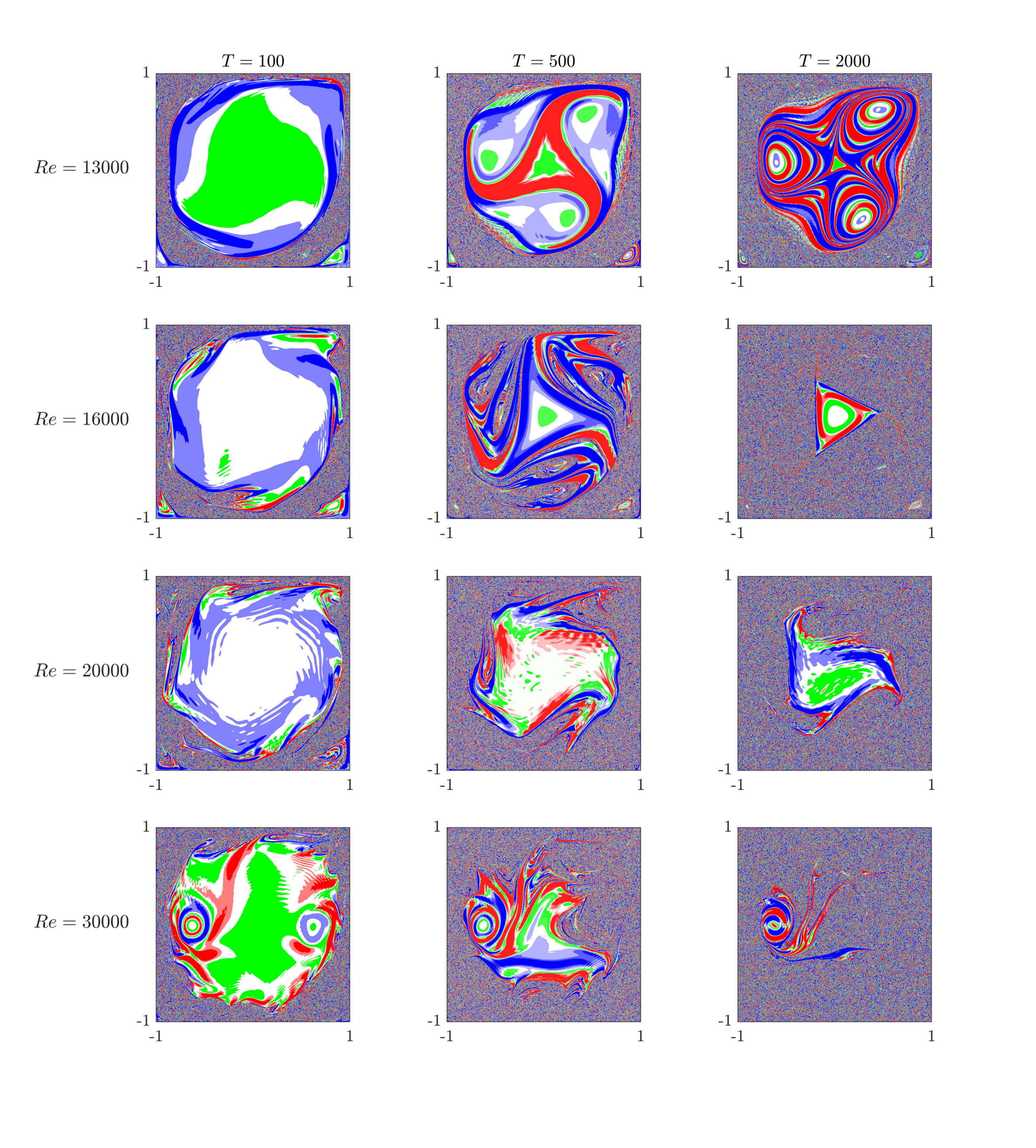}}
	\caption{Hypergraphs of mixing in lid-driven cavity flow on interval $[0,T]$. The zones with combination of red and blue correspond to strong mixing. The figures show that mixing is slower in the center of the cavity.}
	\label{fig:HGs}
\end{figure}

\Cref{fig:HGs} shows the hypergraphs of mixing in cavity flow at different flow regimes and over various time intervals.
The mixing in the cavity flow generally increases as the time-dependence of the velocity field is altered form periodic at $Re\approx 10500$ to fully chaotic at $Re= 30000$, but there is an outstanding feature of mixing  in all those  flows: the mixing in the center of the cavity is slower than mixing in the areas next to the walls. In the periodic flow, the center is dominated by three periodic islands which prevent chaotic mixing. In flows with quasi-periodicity the mixing is stronger but still it leaves a small unmixed  patch in the core.  In the fully chaotic flow, the core gets fully mixed over long intervals but its mixing is still substantially slower than the corner eddies and wall-adjacent areas.
In the following, we explain this phenomenon by revisiting the Prandtl-Batchelor theorem and examining the resonance between the unsteady component of velocity field and circulation time scales of the mean flow.

An important finding of all the previous studies on periodic bounded flows is that the time period of the flow is the most critical parameter that affects the mixing. The experiments by Ottino and co-workers on low-Reynolds lid-driven cavity flows have shown that the time period of the flow governs the existence and size of the periodic islands which determines the strength of mixing   \citep{chien1986laminar,leong1989experiments}.
\citet{ling1992mixing} studied the influence of the flow period on the stability of periodic orbits of tracer motion and detected  different ranges of flow period for which full mixing occurs.
In some studies, the relationship between the time period of the velocity field and the typical tracer circulation period in the steady base flow was exploited to enhance mixing.
\cite{solomon2003uniform}, for instance, performed a numerical and experimental study of a 3D laminar vortex flow and  observed that uniform mixing takes place when the flow period is close to the typical tracer circulation times.
Another related  example is the topological study of mixing in lid-driven cavity flow by \cite{stremler2007generating}. They designed a periodic lid motion that achieves topological chaos by making a delicate match between circulation period of certain tracers and the period of the lid motion.

Here, we use the relationship between the time period of the flow and circulation period of tracers in the \textit{mean flow} to show why mixing in the core of the cavity is slow.
First note that in contrast to the aforementioned studies, the time dependence of cavity flow in our work is due to the flow bifurcations at high Reynolds numbers. Thus the bulk of flow energy is stored in the mean flow and the unsteady component can be considered as a small perturbation to the mean flow. This is evident from the distribution of energy in the Koopman spectrum of the flow in  \cref{fig:spectrum}.
Next, note that the mean flow has a critical feature which is persistent over the considered range of Reynolds number, and that feature is the relatively uniform distribution of vorticity in the core (\cref{fig:Kmodes}). The occurrence of this constant-vorticity core in \textit{steady} flows at high Reynolds numbers is explained by the Prandtl-Batchelor theory. The classical version of this theory states that in regions of the flow with closed streamlines and small viscous forces, the vorticity is constant \citep{prandtl1904uber,batchelor1956steady}.
In a previous work, we showed that this theory also stands for quasi-periodic flows and probably for stationary chaotic flows as well \citep{arbabi2019prandtl}. 
The unsteady version of the theory holds for the central vortex of cavity at high Reynolds, and since the location of this vortex is almost fixed with time,  the vorticity in the core of cavity \emph{mean flow} is uniformly distributed as well.

\begin{figure}
	\centerline{\includegraphics[width=1\textwidth]{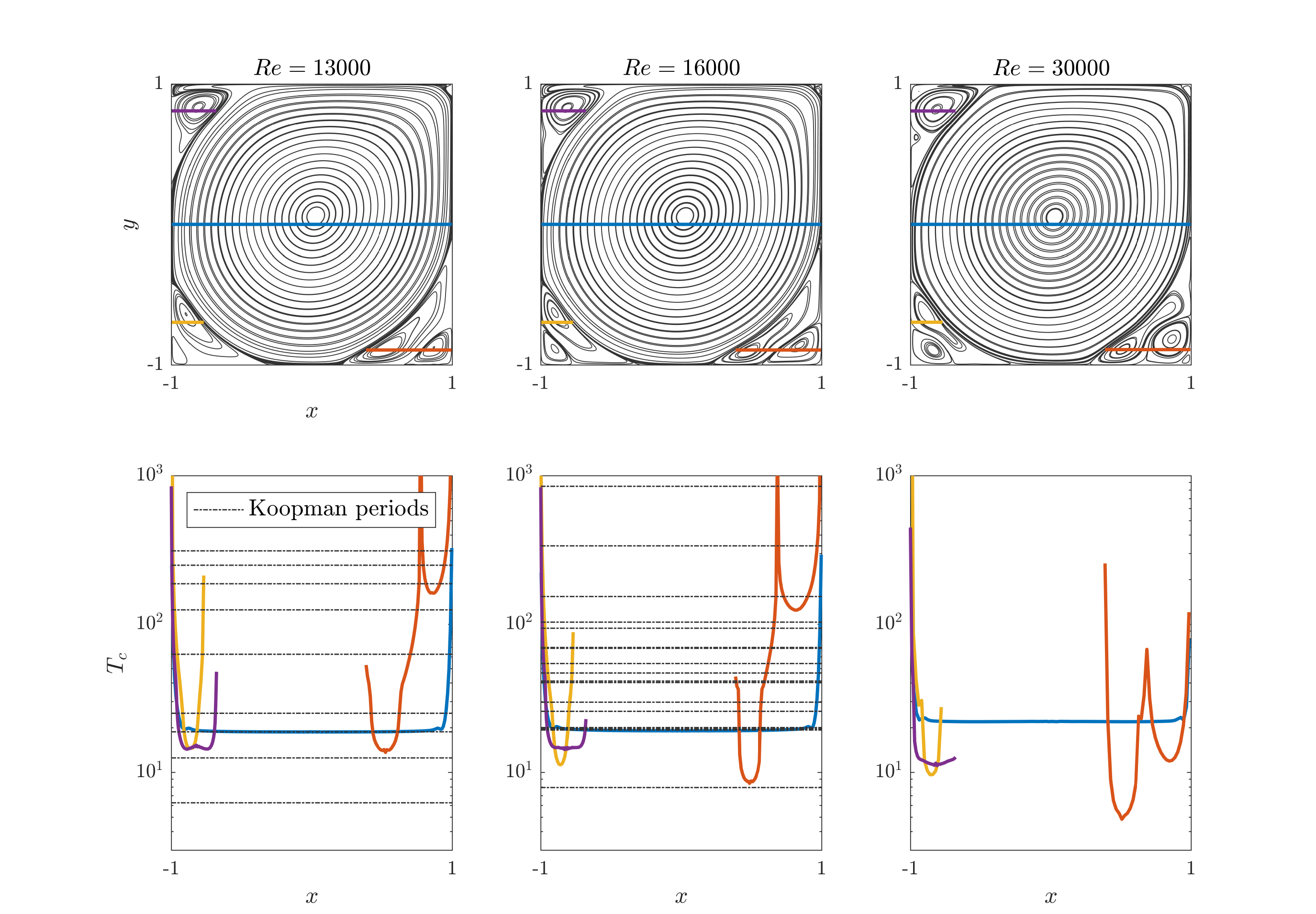}}
	\caption{{ Resonance analysis for mixing: streamlines and initial location of tracers in the mean flow (top), time period of tracer motions and Koopman time periods (bottom). Blue curve in a bottom panel shows the circulation period of tracers released on the blue line in the top panel, and same for other colors. For periodic and quasi-periodic flows only a subset of Koopman periods are shown. The circulation period in the core of all three mean flows is flat which makes resonance less likely.}}
	\label{fig:resonance}
\end{figure}

In the next step,  we show that in the mean flow, the uniform vorticity and closed (but not exactly circular) streamlines lead to a uniform distribution of circulation periods for Lagrangian tracers, that is, the Lagrangian motion is similar to the kinematics of the rigid body rotation.
We consider the nested streamlines in center of the mean flow as shown in any top panel of \cref{fig:resonance}. Let $$p(\psi_0)=\oint_{\psi_0}ds$$ be the perimeter of the streamline that is the $\psi_0$-level set of the mean stream function. We define 
\begin{equation}
  \overline{f}(\psi_0)= \frac{\oint_{\psi_0} f(s)ds}{p(\psi_0)},
\end{equation}
to be the average of function $f$ around that streamline. We can derive a relationship between the average of velocity magnitude on this streamline and vorticity, 
\begin{equation}
  \overline{u}(\psi_0)= \frac{\Gamma(\psi_0)}{p(\psi_0)}= \frac{\int_{A(\psi_0)}\omega dA}{p(\psi_0)}=\frac{\omega_{0}A(\psi_0)}{p(\psi_0)}
\end{equation}
where $\Gamma(\psi_0)$ is the circulation, and $A(\psi_0)$ is the area enclosed by the streamline. In the second equality we have used the Stokes theorem, and in the third we have used constancy of mean vorticity from the Prandtl-Batchelor theorem. 
On the other hand, and given that the velocity on the streamline does not vanish, the time period of circulation for a Lagrangian tracer around the streamline is given by
\begin{equation}\label{eq:PBT0}
  T_c(\psi_0)= \oint_{\psi_0}\frac{1}{u(s)}ds= \overline{\bigg(\frac{1}{u}\bigg)}(\psi_0)p(\psi_0).
\end{equation}
Here we make the assumption that variations of velocity around the streamline is small compared to its average (e.g. the streamline is close to circular shape), that is, $u(s)=\overline{u}+u'(s)$ with $|u'(s)|\ll\overline{u}$, so that
\begin{equation}
  \overline{\bigg(\frac{1}{u}\bigg)}=\overline{\bigg(\frac{1}{\overline{u}+u'}\bigg)}=\frac{1}{\overline{u}}\overline{\bigg(\frac{1}{1+u'/\overline{u}}\bigg)}\approx \frac{1}{\overline{u}}\overline{(1-u'/\overline{u})}=\frac{1}{\overline{u}}.
\end{equation}
Using this approximation in \eqref{eq:PBT0}, we get
\begin{equation}\label{eq:PBT}
  T_c(\psi) \approx \frac{1}{\overline{u}(\psi_0)}p(\psi_0)=\frac{p^{2}(\psi_0)}{\omega_{0}A(\psi_0)}=\frac{s(\psi_0)}{\omega_{0}}.
\end{equation}
where we have defined $s:=p^{2}/A$. The parameter $s$ is a property of the shape of the streamline and therefore two similar-looking streamlines would have the same circulation period regardless of their sizes. 
In particular, the streamlines in the core of the mean cavity flow (\cref{fig:resonance} top row) are quite similar which leads to almost uniform distribution of circulation period.   As shown in bottom row of  \cref{fig:resonance}, the numerical computation of those time periods confirms this analysis (the blue curves). In contrast, the circulation periods of the smaller vortices  vary largely over shorter lengths (red, yellow and cyan curves). This observation also holds for the quasi-periodic and stationary chaotic flow due to their mean flow structure.

{
For the special case of periodic flow, the slow mixing in the core is explained by the classical perturbation analysis of Hamiltonian systems. We consider the motion of tracers in the mean flow to be a 2D dynamical system, and the mean stream function to be its Hamiltonian. The unsteady component of the flow serves as a time-periodic perturbation to the vector field of this Hamiltonian system. Then the question of whether the flow is well-mixed and where the mixing occurs translates to identifying the size and location of chaotic regions in the perturbed Hamiltonian system.
There are a few techniques such as Melnikov method or Kolmogorov-Arnold-Moser (KAM) theory to predict whether any chaotic trajectories appear in such a system  \citep{guckenheimer1983nonlinear}. 
The essence of these techniques is to detect the \emph{resonances} between the circulating trajectories of the Hamiltonian system (i.e. the mean flow) and the perturbation (i.e. unsteady component of the fluid flow) which leads to chaotic motion.
Now if we contrast the circulation periods of the mean flow at $\Rey=13000$ to the time period of the flow and its harmonics (i.e. Koopman time periods, shown in the bottom left panel of fig. \ref{fig:resonance}), we see that the motion of tracers in the corner eddies allow many resonances with the perturbing flow field, and hence, those regions contain a larger number of chaotic trajectories, whereas the flat distribution of time periods in the central vortex would not allow so many chances for resonance and  the amount of chaotic trajectories in the center would be substantially lower.
\Cref{fig:HGs}, indeed, shows that regions of secondary corner vortices undergo substantial mixing while the core is dominated by regular trajectories.
}

\subsection{Hypothesis testing for the effect of Prandtl-Batchelor theorem on mixing}
\begin{figure}
	\centerline{\includegraphics[width=1.0\textwidth]{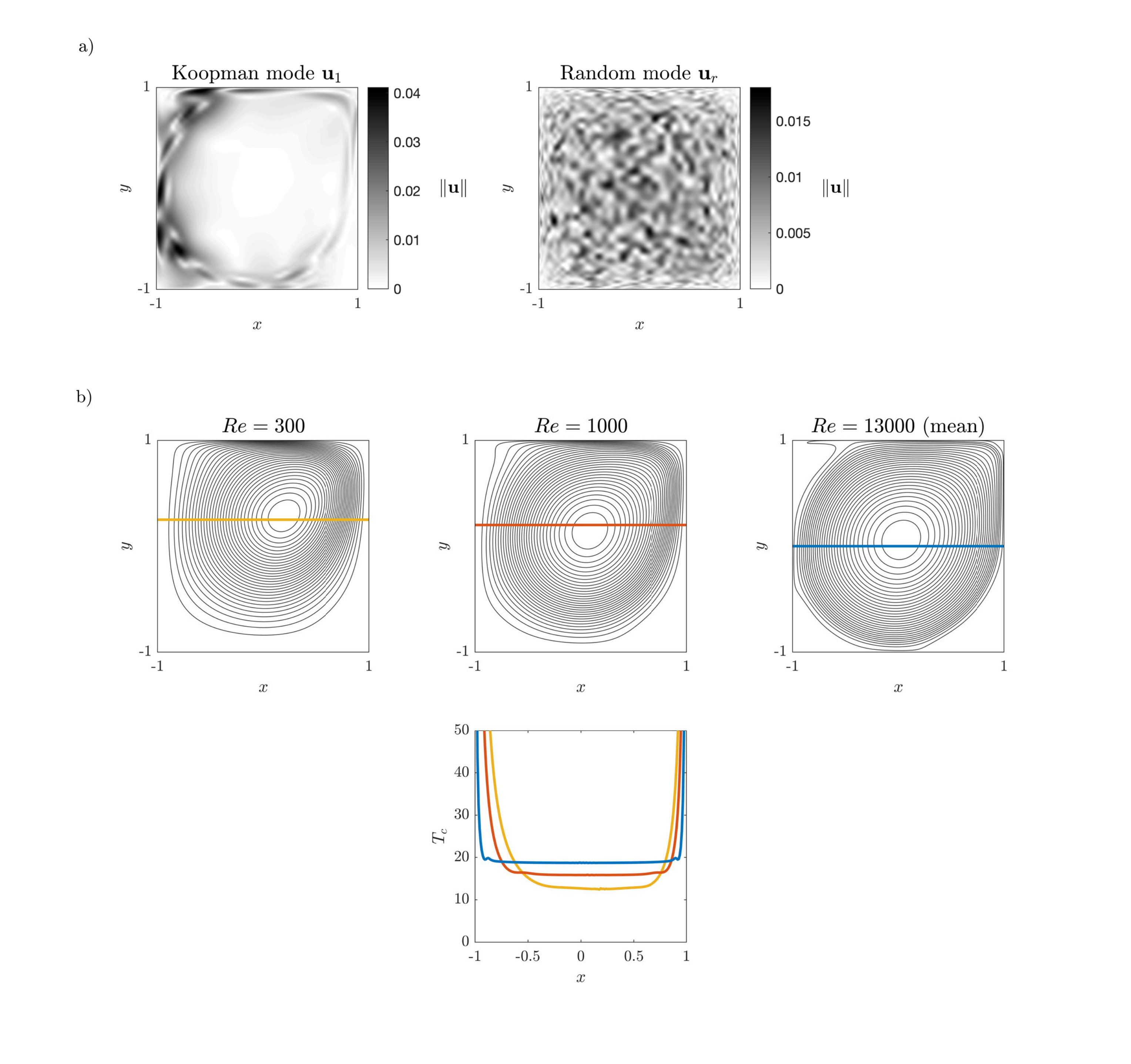}}
	\caption{The setup for testing the effect of vorticity distribution on mixing: a) magnitude of velocity in the first oscillatory Koopman mode of $Re=13000$ and a randomly generated velocity field with the same kinetic energy, b) top row: streamlines of central vortex in steady flows at $Re=300,1000$ and the mean flow at $Re=13000$, bottom row: the tracer period distributions in those flows. We use the random field to perturb the steady flows and disambiguate the effect of vorticity distribution from the special structure of oscillatory Koopman modes. $T^*$ is the time normalized by core rotation periods.
  }
	\label{fig:perturb}
\end{figure}

An alternative hypothesis for explanation of slow mixing in the core would be to note that the unsteady component of velocity field is weaker in the core compared to edges of the central vortex. Indeed, \cref{fig:perturb}(a) shows that in the first oscillatory Koopman mode, the kinetic energy is mostly distributed along the edge of the central vortex. In order to disambiguate and emphasize  the effect of constant vorticity due to Prandtl-Batchelor theorem on the mixing we conduct a numerical experiment as follows:

We first generate a random mode of velocity field $\mathbf{u}_r(x,y)$ on the cavity domain. To this end,  we create a random stream function field $\psi_r(x,y)=(1-x^4)(1-y^4)q(x,y)$ where $q$ is generated by drawing a random value from standard normal distribution for each computational grid point, and then set 
\begin{equation}\label{eq:u_r}
 \mathbf{u}_r= \big[ \frac{\p \psi_r }{\p y}, ~-\frac{\p \psi_r }{\p x}\big]^\top
\end{equation}
We also scale $\mathbf{u}_r$ such that its kinetic energy is equal to that of the first oscillatory Koopman mode at $Re=13000$. This random field is incompressible, stronger in the core compared to the wall-adjacent areas and satisfies the no-slip boundary condition (\cref{fig:perturb}(a)).

We use the random field $\mathbf{u}_r$ to perturb the velocity fields with various types of vorticity distribution and study the mixing in those perturbed flows. Consider the artificial flow model given as
\begin{equation}  \label{eq:art_model}
\mathbf{u} = \mathbf{u}_b + \alpha \mathbf{u}_re^{i\beta t}.
\end{equation}
In the first model, we choose the base flow $\mathbf{u}_b $ to be the mean of periodic flow at $Re=13000$, and set $\alpha=1,\beta=\omega_1$ where $\omega_1$ is the Koopman basic frequency of the periodic flow. This model is like the actual periodic flow at $Re=13000$, except that we have replaced the time-dependent component with a spatially random mode which oscillates at the same frequency with a similar amount of kinetic energy. 
In the next two models, as the base flow we use the steady cavity flows at $Re=300$ and $Re=1000$. As shown in \cref{fig:perturb}(b), these two flows also have rotating cores but the tracer period distribution in their core is not as flat as the mean flow at  $Re=13000$. 
To put these two models on the same footing as the first model, we choose $\alpha$ such that the ratio $\alpha\|\mathbf{u}_r\|/\|\mathbf{u}_b\|$ is the same for all three models. Also, we choose $\beta$ such that the ratio between the time period of perturbation and the rotation time for the core is the same across the models. To be more precise, let $T_c$ be the minimum circulation period at the core of the base flow. Then for all the models, we choose  
\begin{equation}  \label{eq:art_model}
\beta = \omega_1 \frac{T_c^{Re=13000}}{T_c}.
\end{equation}
Next we compute the hypergraphs of these three models over windows of various lengths. To remove the effect of time on quality of mixing, we normalize the length of those windows by scaling the time with respect to the core rotation time, that is, we use
\begin{equation}  \label{eq:art_model}
T^* = \frac{T}{T_c}.
\end{equation}

\begin{figure}
	\centerline{\includegraphics[width=1.0\textwidth]{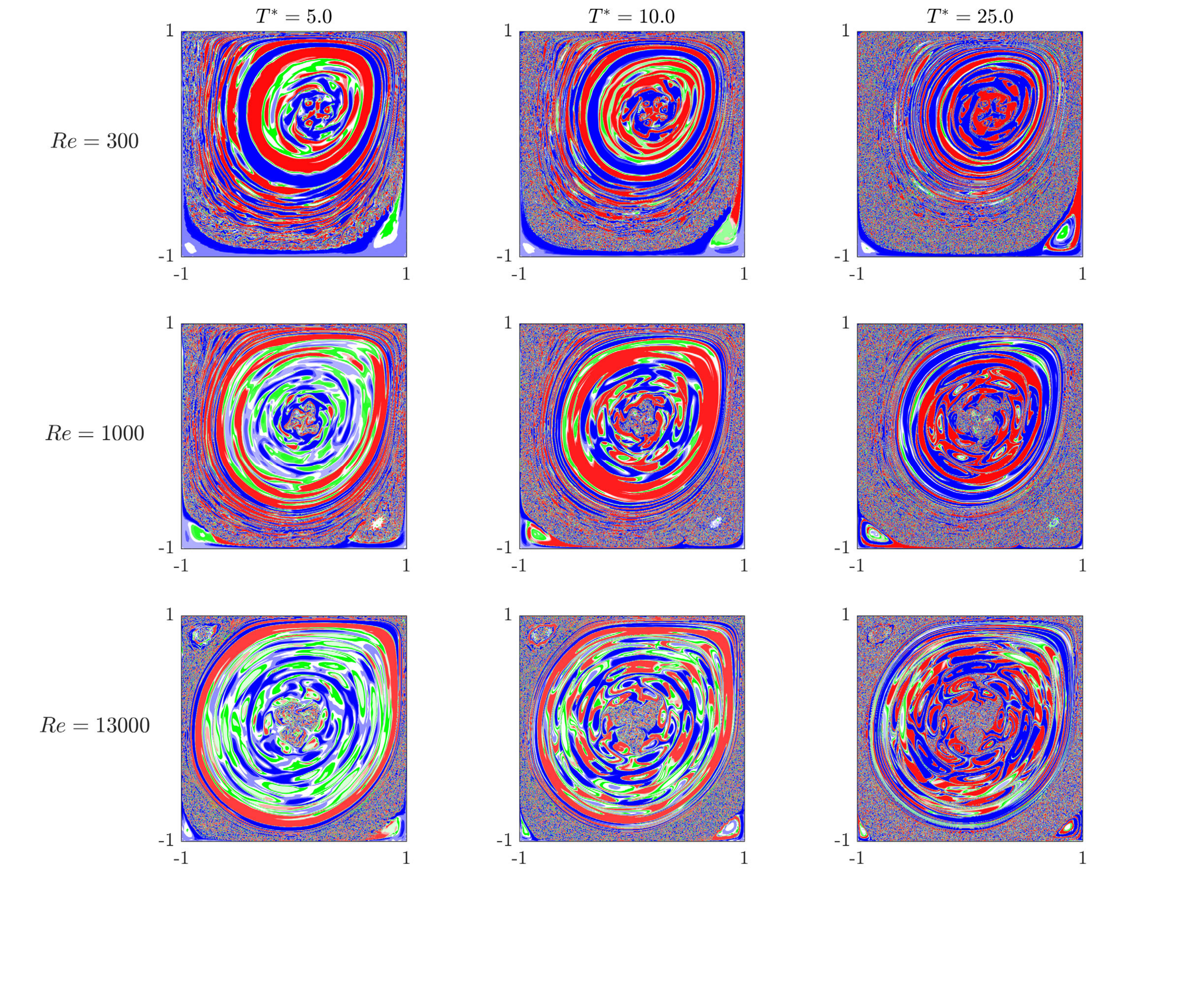}}
	\caption{Effect of Prandtl-Batchelor theorem on mixing: hypergraphs of mean flows at various Reynolds numbers perturbed by the oscillating random field in \cref{fig:perturb}(a). Higher Reynolds number leads to a more uniform distribution of circulation periods in the core of mean flow and more resilience against perturbation, and hence, weaker mixing.}
	\label{fig:perturb2}
\end{figure}

The results of the above experiment, shown in \cref{fig:perturb2}, indicate that flatness of the circulation period makes the core of the flow more resilient to perturbations, that is, the core with a flatter distribution takes longer to mix.  Therefore, it confirms the adverse effect of Prandtl-Batchelor theorem on mixing at high Reynolds numbers and regardless of some  specific features of the oscillating mode. Another way to interpret this phenomenon is to note that although shear between the rotating layers of a steady flow is not enough to generate chaotic mixing, but higher shear exposed to time-dependent perturbations leads to stronger mixing.  On the other hand, the Prandtl-Batchelor theorem dictates a nearly shearless core for the mean of rotational flows at  high Reynolds, and hence predicts that mixing in that core is slower than other areas exposed to shear such as areas adjacent to the walls.  

Finally, we note that although the bulk of the core remains unmixed in \cref{fig:perturb2}, a well-mixed patch emerges at the very center. This is an artifact of our experiment: The randomly generated field is much stronger than base flow at the very center and prediction of mixing with perturbation argument is not correct in that region.  

\subsection{Quasi-periodic and aperiodic flows}
\begin{figure}
	\centerline{\includegraphics[width=1 \textwidth]{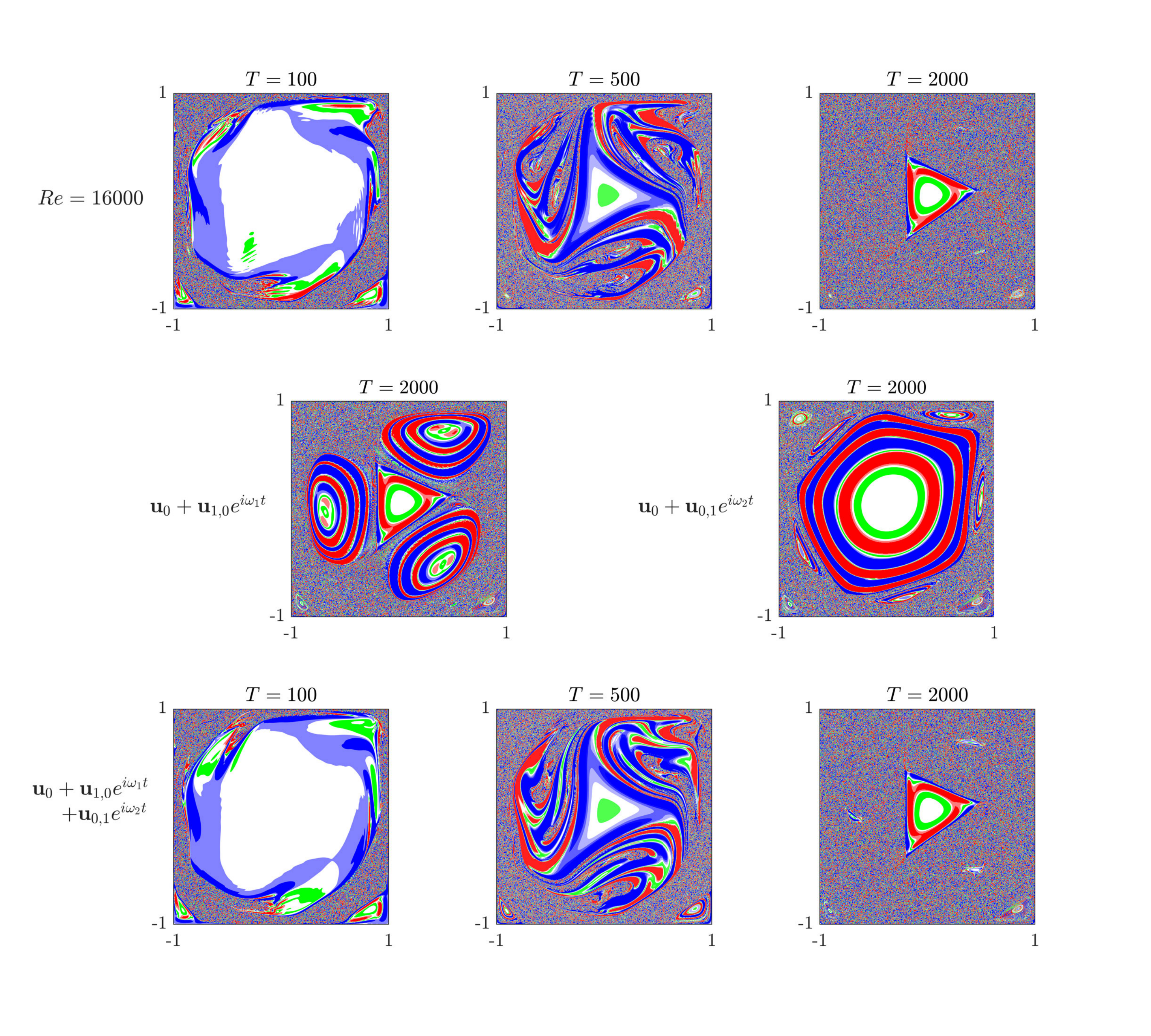}}
	\caption{Hypergraphs of the quasi-periodic flow repeated from \cref{fig:HGs} (top), hypergraphs of flow models each consisting of the mean flow and one basic Koopman mode (middle), and hypergraphs of a flow model consisting of the mean flow and both basic Koopman modes (bottom). The resonances between the two oscillatory modes leads to stronger mixing in the core compared to periodic flow but still slower compared to wall-adjacent areas. }
	\label{fig:HGRe16}
\end{figure}

{
Our qualitative argument based on variation of the Lagrangian circulation periods and chance of resonance with time-dependent perturbations can be applied to quasi-periodic and aperiodic flows as well.  In the former case, if we synthesize a new flow which consists of the mean flow and only one of basic oscillating Koopman modes, then we observe that mixing is much weaker  in the core of the flow, and depending on the Koopman mode and frequency a different number of periodic islands will appear (middle row of \cref{fig:HGRe16}). This is while the secondary vortices undergo much faster mixing similar to the periodic case. Because the frequencies in a quasi-periodic flow are incommensurate, their combined harmonics (i.e. linear combination of the two basic frequencies with integer coefficients) are dense in any frequency interval. 
Therefore when we only add the two basic oscillatory modes to the mean flow, the chance of resonance with the circulation periods of the mean flow dramatically increases which leads to a larger amount of chaotic trajectories.
This results in a core mixing which is stronger than the periodic flow but still weaker than the secondary vortices and wall-adjacent areas, because the variation of the circulation periods in the core is still much smaller than corner eddies (middle column in \cref{fig:resonance}).

Application of this idea to chaotic flows requires a standard observation from dynamical systems theory: 
The evolution of stationary chaotic systems over any finite time interval can be approximated with arbitrary accuracy using a quasi-periodic evolution \citep[see e.g.][]{korda2018data}. Therefore, the finite-time mixing of the stationary chaotic flow resembles a quasi-periodic model with sufficiently many oscillatory modes. 
The number of modes and frequencies required to approximate the chaotic component are typically large which leads to generally faster mixing, but again the mixing in core is slower than other areas due to uniform distribution of vorticity and circulation periods. This explains the picture of mixing at $Re=30000$ in  \cref{fig:HGs}.  
}

The resilience of rotational cores toward mixing at high Reynolds  numbers has been already observed in the context of geophysical flows. \cite{geyer2010mixing} have measured mixing in an estuary where a fresh water plume from the river is flowing onto the saline water from the ocean. The measurements show that the cores of Kelvin-Helmholtz vortices formed at the interface of the fluid layers exhibit poor mixing and the bulk of mixing is occurring at the edge of those vortices and the thin filaments that connect them.  They contrast their result with the previous numerical and experimental studies which are carried out at moderate and low Reynolds numbers and conclude that the weak mixing in the core is a characteristic of high-Reynolds flows.
In their study this behavior is attributed to the shear instabilities in the edge of vortices and connecting filaments. 
The analysis in the above paragraphs supports this characterization from a kinematic standpoint, namely, the Prandtl-Batchelor theory of unsteady flows predicts an almost shearless core at high Reynolds, and the ensuing rigid-body motion of tracers in the core makes it more resilient toward advective mixing when it is exposed to small-amplitude unsteady perturbations. We acknowledge that our arguments are based on analysis of homogenous 2D flows, and therefore their extension to 3D stratified flows requires further study.

\section{Approximation of mixing using projected models} \label{sec:approx}
{
We argued in the previous section that with sufficiently many oscillatory modes one can approximate the mixing in aperiodic flows over finite time.  We also isolated the study of periodic and quasi-periodic flow to their dominant modes. However,
approximating the Lagrangian properties of a flow with a truncated modal expansion could be problematic because small errors in representation of velocity field can lead to exponentially growing errors in estimation of tracer trajectories. 
The main purpose of this section is to verify that modal approximations, such as the ones used in the previous section, are reliable for understanding the general picture of mixing. Therefore, instead of looking at individual tracer trajectories, which correspond to the \emph{microscopic} picture of mixing, we look at the error in the advection of concentration fields in the flow.
This type of analysis could also be helpful in determining the suitable truncation dimension or appropriate level of ``denoising''  in  data reconstruction methods \citep[e.g.][]{bui2004aerodynamic} and model reduction techniques that rely on projected models \citep[e.g.][]{cazemier1998,noack2003hierarchy}.
}


We use Koopman modes and POD modes in our analysis, mainly because the Koopman and POD modes are the most economical choices for constructing low-dimensional approximations of the quasi-periodic and chaotic components of the flow respectively \citep{holmes2012turbulence,arbabi2017study}. 
We first construct projected models of the flow which contain only a subset of Koopman and POD modes.
Given the definition and notation of KMD and POD in \cref{sec:cavityKMD}, we define a projected model as
\begin{equation}  \label{eq:proj_model}
\mathbf{u}_{n_k,n_p}(\mathbf{x},t)=  \sum_{j=1}^{n_k} \mathbf{u}_j(\mathbf{x}) e^{i\omega_j t} + \sum_{l=1}^{n_p}  \boldsymbol{\phi}_l(\mathbf{x})a_l(t),
\end{equation}
where $n_k$ and $n_p$ denote the number of Koopman and POD modes used in the model. In forming the above expansion, we stack the modes by starting from the Koopman modes (including the mean flow) in the order of decreasing energy and then use the POD modes of the chaotic component with the same ordering.

\begin{figure}
	\centerline{\includegraphics[width=1\textwidth]{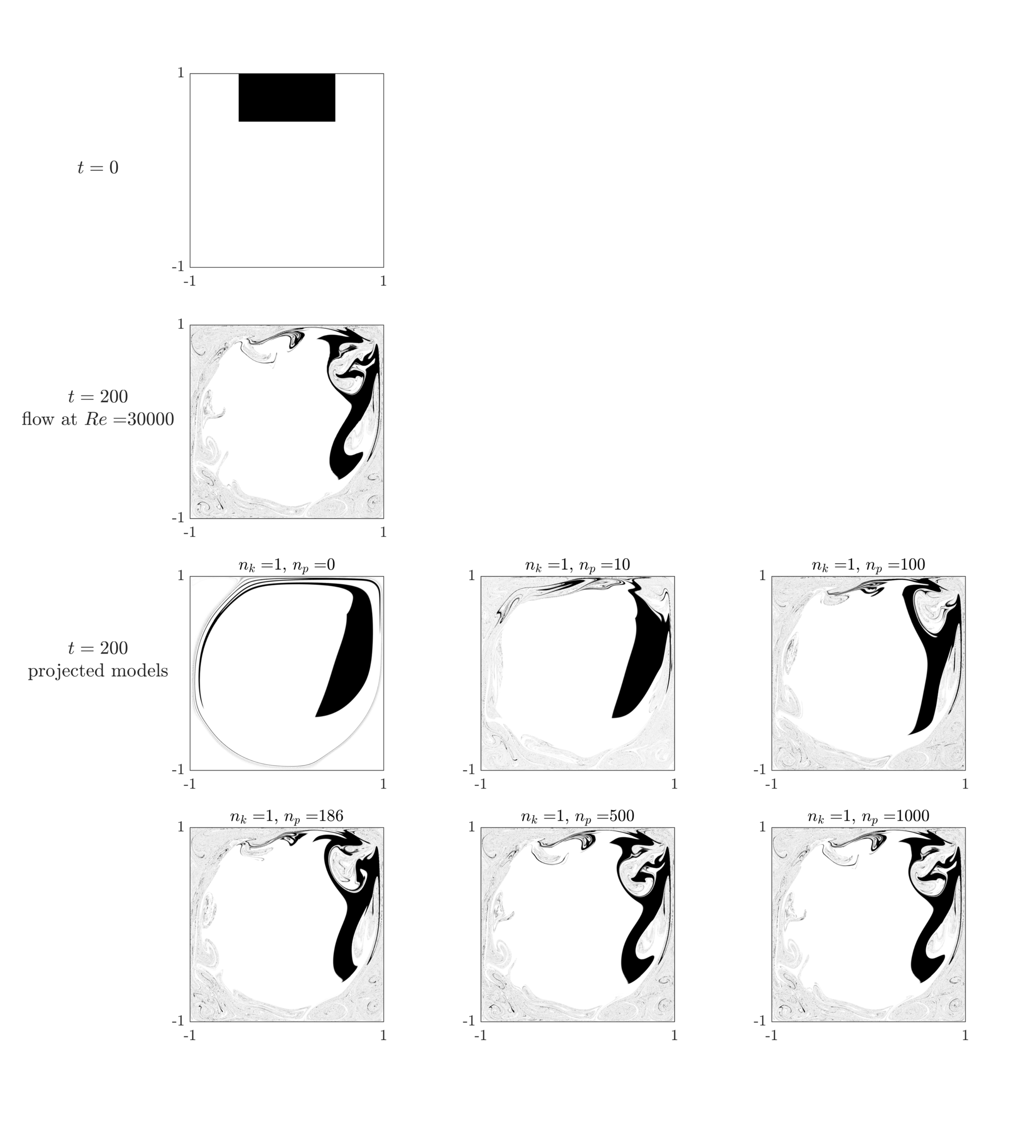}}
	\caption{The initial concentration field (first row), concentration field advected by the chaotic flow (second row), and concentration field advected by the projected models defined in \eqref{eq:proj_model} (bottom rows). $n_k$ and $n_p$ are, respectively, the number of Koopman and POD modes in the projected model.  }
	\label{fig:mixing19}
\end{figure}

We compare the mixing in the flow and its projected model using the following numerical experiment: we define an initial blob of concentration as shown on top of \cref{fig:mixing19}. We advect this blob in the actual flow as well as the projected models. Then over regular intervals we measure how much the advected concentration field in the projected flows is deviating from the actual flow by defining
 \begin{equation}\label{eq:mixerror}
   {e}(t)=\frac{\Phi(\Delta c(\mathbf{x},t))}{\Phi(c(\mathbf{x},0))}
 \end{equation}
where $c(\Delta \mathbf{x},t)$ is the difference of the concentration fields, $c(\mathbf{x},0)$ is the initial concentration field corresponding to the rectangular blob and  $\Phi(\cdot)$ is the mix-norm  defined in \eqref{eq:MixNorm}. Note that the results of this experiment depends on the initial condition of the blob, however, through several trials, we have chosen the initial position so as to incorporate mixing near and away from the walls (\cref{fig:mixing19}) and give a clear and general picture of mixing.

\begin{figure}
	\centerline{\includegraphics[width=1 \textwidth]{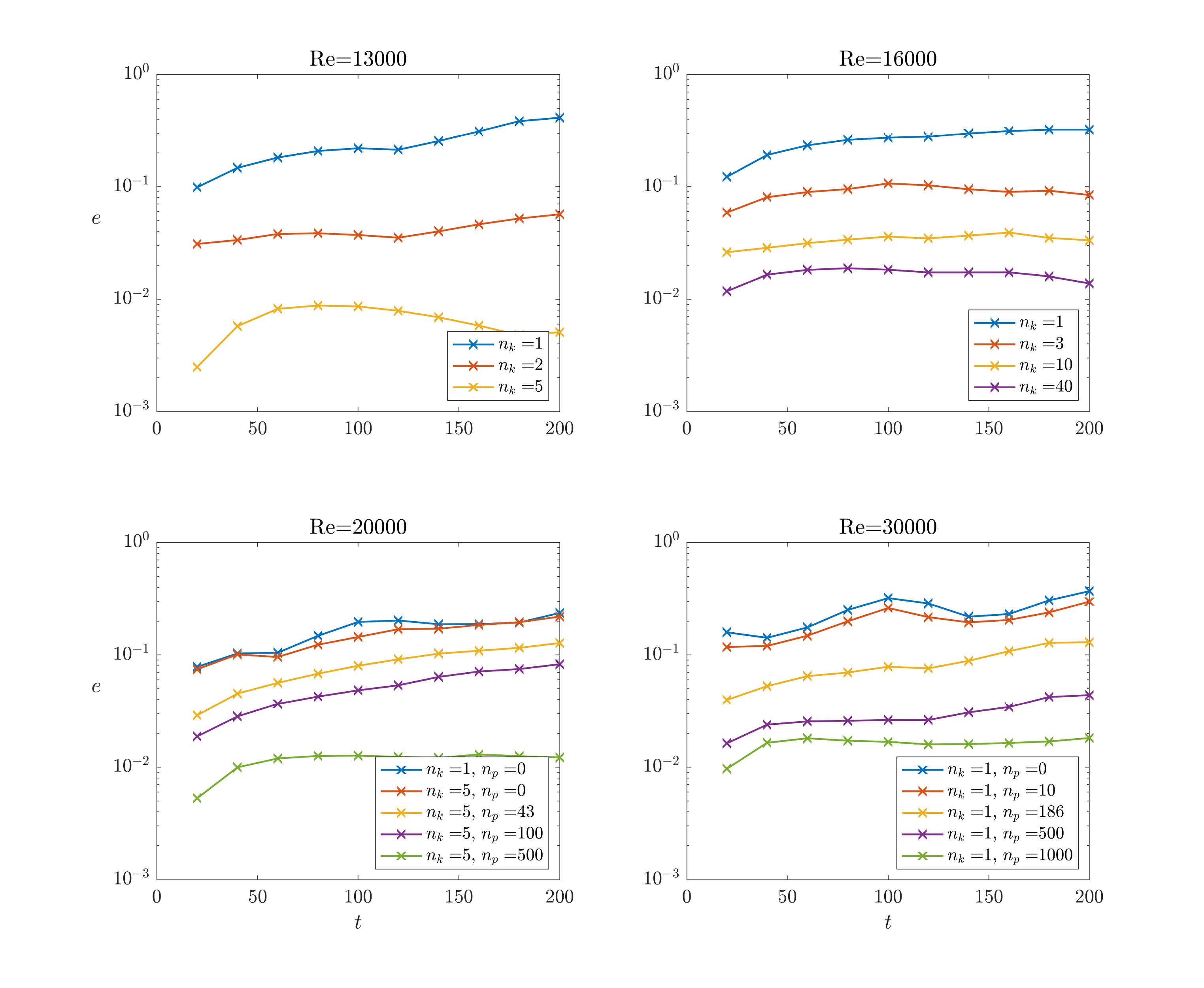}}
	\caption{Mix-norm error of mixing approximation using projected models. $n_k$ and $n_p$ denote the number of Koopman modes and POD modes in a projected model. The projected model with $n_p=43$ and $n_p=186$ resolve $99\%$ of the unsteady kinetic energy in the flow at $Re=20000$ and $Re=30000$ respectively. }
	\label{fig:mixerror}
\end{figure}

 \Cref{fig:mixerror} shows the results of the experiment over an interval of length 200 which is about 10  times the circulation period in the core of the mean flow. 
 At the start the deviation rapidly grows  from zero due to difference in velocity fields (this growth is not shown in the log-log plot) but as the time passes, the rate of error growth slows down because the mixing process transfers more concentration to small scales which are suppressed in the mix-norm.
 The mixing error shows that the periodic flow has the lowest complexity in the sense that two Koopman modes (i.e. mean flow + basic oscillatory mode) are enough to describe the advection of the blob within an accuracy of a few percents over long times. A similar observation holds for the quasi-periodic flow where using 3 modes (mean flow and the two basic oscillatory modes) recovers the mixing with more than 90\% accuracy. Hence, one could restrict the quantitative  analysis, e.g. using perturbation methods in chaotic advection, to these few modes and obtain a reasonable approximation of transport rates and other quantities of interest related to mixing.
 
 {
The small error in the velocity field of a projected model typically leads to exponential divergence of tracer trajectories from the ones in the actual flow. An interesting outcome of our experiment is that even with this exponential growth of trajectory errors, the approximation of error in the mix-norm, which is a macroscopic measure of mixing, has a  gentle growth.  For example,  in chaotic flows at $Re=20000$ and $Re=30000$, the projected models that contain $99\%$ of the unsteady kinetic energy can approximate mixing with $2-4\%$ error after one circulation period, and 10\% over the whole time interval (yellow curves in the bottom row of \cref{fig:mixerror}).  
Although the number of modes required for accurate prediction of mixing drastically increases for chaotic flows, projected models still offer a scalable approximation of the mixing in the original flow. }

\section{Conclusion} \label{sec:conclusion}
An interesting feature of 2D cavity flow at high Reynolds numbers is that mixing in its core is weaker than corner eddies and wall-adjacent areas. In this paper, we explained this observation through a combination of ideas from fluid dynamics and dynamical systems. Namely, we showed that the circulation periods in the mean flow have a flat distribution in the core as a consequence of Prandtl-Batchelor theorem, and the poor mixing is the result of the interaction between this flat distribution and the time scales of Eulerian flow perturbations, that is, unsteady components of the flow which possess small amplitudes compared to the mean flow. This qualitative analysis also holds for quasi-periodic and chaotic flows. 
These results support the field observations in stratified flows that show poor mixing in the core of Kelvin-Helmholtz vortices at high Reynolds.

We also characterized the contribution of Koopman and POD modes to mixing at various Reynolds numbers by comparing the advection of a generic blob in the actual flow versus projected models of flow consisting of a subset of those modes. Our analysis showed that although individual trajectories maybe approximated with an exponentially growing error, the 
mix-norm error of the advected field does not grow exponentially fast. As a result, a sufficiently large proper subset of modes can be used to approximate the macroscopic picture of mixing over finite time intervals with desired accuracy.

\section{Acknowledgement and source code}
This research was partially supported by the ONR grant N00014-14-1-0633. H. A. thanks Pierre-Yves Passaggia for an informative discussion and introducing references on mixing in stratified flows and  S. Mohammad Mirzadeh for helpful notes on advection simulations. We are also grateful to L. Gary Leal for comments and questions that improved this manuscript.

The MATLAB implementation of advection, mix-norm and hypergraphs computation along with the flow data used in the paper is available at \url{https://github.com/arbabiha/Mixing-analysis-2d-flows}.

\section*{Appendix: computation and visualization of hypergraphs}\label{Appendix}

 Consider the trajectory of a passive tracer starting at $\mathbf{x}$ at time $t_0$. The position of this tracer at time $t>t_0$ is given by the flow map $\mathbf{F}(\mathbf{x},t)$, which solves the ordinary differential equation (ODE)
\begin{equation}
\dot{\mathbf{F}}(\mathbf{x},t) = \mathbf{u}(\mathbf{F}(\mathbf{x},t),t), \quad t\in[t_0,t_0+T],\quad \mathbf{F}(\mathbf{x},t_0)=\mathbf{x}, \label{eq:advection}
\end{equation}
where $\mathbf{u}$ denotes the velocity field. The average Lagrangian velocity of the tracer is related to the flow map via
 \begin{equation}
 \mathbf{u^*}_{t_0}^{t_0+T}(\mathbf{x})= \frac{1}{T}(\mathbf{F}(\mathbf{x},t_0+T)-\mathbf{x}).
 \end{equation}

Computing the mesohyperbolicity field defined in (\ref{eq:mesohyper}) requires two steps of computation.
First is to solve the ODE in (\ref{eq:advection}). This is often achieved by direct integration for a grid of passive tracers initially distributed over the flow domain.
The second step is to evaluate the gradient of Lagrangian velocity. There are two approaches to accomplish this step.
In the first approach, the gradient is computed via finite difference on the initial grid of tracers.
An auxiliary grid collocated around points of the initial grid can be used to improve the numerical efficiency \citep[see e.g. the computation of the flow map gradient by][]{Farazmand2012}.
In the second approach, which we take here, the use of finite difference is avoided by integrating the variational equation for the flow map gradient \citep{anosov1994dynamical},
 \begin{equation}
  \frac{\mathrm{d}}{\mathrm{d}t}\nabla\mathbf{F}(\mathbf{x},t)=\nabla \mathbf{u}(\mathbf{F}(\mathbf{x},t),t)\cdot\nabla\mathbf{F}(\mathbf{x},t), \quad t\in[t_0,t_0+T], \quad \nabla\mathbf{F}(\mathbf{x},t_0)=I,\label{eq:gradF}
\end{equation}
with $I$ being the identity matrix. This ODE is solved along with \eqref{eq:advection} for the same set of tracers.
The Lagrangian velocity gradient is given by
 \begin{equation}
    \nabla\mathbf{u^*}_{t_0}^{t_0+T}(\mathbf{x})= \frac{1}{T}(\nabla\mathbf{F}(\mathbf{x},t_0+T)-I).
\end{equation}
This approach  requires nearly the same computational effort as the finite difference approach, given that the instantaneous field of $\nabla \mathbf{u}$ is readily available.
We used  the standard 4th-order Runge-Kutta method to integrate equations \eqref{eq:advection} and \eqref{eq:gradF}. In doing so, the velocity field obtained by numerical solution of Navier-Stokes equations was interpolated using the spline method in space, and the linear method in time.
In the hypergraphs plotted in this paper, the mesohyperbolicity field, $M(\mathbf{x}):=\det|\nabla \mathbf{u^*}_{t_0}^{t_0+T}(\mathbf{x})|$, is plotted and partitioned into the mesohyperbolic ($M<0$), mesoelliptic ($0<M<4/T^2$), and mesohelical ($M>4/T^2$) regions, which are respectively marked by blue, green and red colors. For  more readability, the value of mesohyperbolicity field in hypergraphs is cut off for $M>8/T^2$ and $M<-4/T^2$.

\bibliographystyle{jfm}
\bibliography{allMixing.bbl}

\begin{thebibliography}{64}
\expandafter\ifx\csname natexlab\endcsname\relax\def\natexlab#1{#1}\fi
\def\au#1{#1} \def\ed#1{#1} \def\yr#1{#1}\def\at#1{#1}\def\jt#1{\textit{#1}}
  \def\bt#1{#1}\def\bvol#1{\textbf{#1}} \def\vol#1{#1} \def\pg#1{#1}
  \def\publ#1{#1}\def\arxiv#1{#1}\def\org#1{#1}\def\st#1{\textit{#1}}

\bibitem[Anderson {\em et~al.\/}(1999)Anderson, Galaktionov, Peters, Van~de
  Vosse \& Meijer]{anderson1999analysis}
{\sc \au{Anderson, PD}, \au{Galaktionov, OS}, \au{Peters, GWM}, \au{Van~de
  Vosse, FN} \& \au{Meijer, HEH}} \yr{1999}  \at{Analysis of mixing in
  three-dimensional time-periodic cavity flows}.  \jt{Journal of Fluid
  Mechanics}  \bvol{386},  \pg{149--166}.

\bibitem[Anderson {\em et~al.\/}(2000)Anderson, Galaktionov, Peters, van~de
  Vosse \& Meijer]{anderson2000chaotic}
{\sc \au{Anderson, Patrick~D}, \au{Galaktionov, Oleksiy~S}, \au{Peters,
  Gerrit~WM}, \au{van~de Vosse, Frans~N} \& \au{Meijer, Han~EH}} \yr{2000}
  \at{Chaotic fluid mixing in non-quasi-static time-periodic cavity flows}.
  \jt{International journal of heat and fluid flow}  \bvol{21}~(2),
  \pg{176--185}.

\bibitem[Anosov \& Arnold(1994)]{anosov1994dynamical}
{\sc \au{Anosov, DV} \& \au{Arnold, VI}} \yr{1994} {\em Dynamical System I,
  Ordinary Differential Equations and Smooth Dynamical Systems\/}.
  \publ{Springer Verlag}.

\bibitem[Arbabi \& Mezi\'{c}(2017)]{arbabi2017study}
{\sc \au{Arbabi, Hassan} \& \au{Mezi\'{c}, Igor}} \yr{2017}  \at{Study of
  dynamics in post-transient flows using {Koopman} mode decomposition}.
  \jt{Phys. Rev. Fluids}  \bvol{2},  \pg{124402}.

\bibitem[Arbabi \& Mezi{\'c}(2019)]{arbabi2019prandtl}
{\sc \au{Arbabi, Hassan} \& \au{Mezi{\'c}, Igor}} \yr{2019}
  \at{Prandtl--batchelor theorem for flows with quasiperiodic time dependence}.
   \jt{Journal of Fluid Mechanics}  \bvol{862}.

\bibitem[Aref(1984)]{aref1984}
{\sc \au{Aref, Hassan}} \yr{1984}  \at{Stirring by chaotic advection}.
  \jt{Journal of fluid mechanics}  \bvol{143},  \pg{1--21}.

\bibitem[Batchelor(1956)]{batchelor1956steady}
{\sc \au{Batchelor, G~K}} \yr{1956}  \at{On steady laminar flow with closed
  streamlines at large reynolds number}.  \jt{Journal of Fluid Mechanics}
  \bvol{1}~(02),  \pg{177--190}.

\bibitem[Boyland {\em et~al.\/}(2000)Boyland, Aref \&
  Stremler]{boyland2000topological}
{\sc \au{Boyland, Philip~L}, \au{Aref, Hassan} \& \au{Stremler, Mark~A}}
  \yr{2000}  \at{Topological fluid mechanics of stirring}.  \jt{Journal of
  Fluid Mechanics}  \bvol{403},  \pg{277--304}.

\bibitem[Budi{\v{s}}i{\'c} \& Thiffeault(2015)]{budisic2015finite}
{\sc \au{Budi{\v{s}}i{\'c}, Marko} \& \au{Thiffeault, Jean-Luc}} \yr{2015}
  \at{Finite-time braiding exponents}.  \jt{Chaos: An Interdisciplinary Journal
  of Nonlinear Science}  \bvol{25}~(8),  \pg{087407}.

\bibitem[Budi\v{s}i{\'c} {\em et~al.\/}(2016)Budi\v{s}i{\'c}, Siegmund, Son \&
  Mezi{\'c}]{Budisic2016c}
{\sc \au{Budi\v{s}i{\'c}, Marko}, \au{Siegmund, Stefan}, \au{Son, Doan~Thai} \&
  \au{Mezi{\'c}, Igor}} \yr{2016}  \at{Mesochronic classification of
  trajectories in incompressible {{3D}} vector fields over finite times}.
  \jt{Discrete and Continuous Dynamical Systems - Series S}  \bvol{9}~(4),
  \pg{923--958}.

\bibitem[Bui-Thanh {\em et~al.\/}(2004)Bui-Thanh, Damodaran \&
  Willcox]{bui2004aerodynamic}
{\sc \au{Bui-Thanh, Tan}, \au{Damodaran, Murali} \& \au{Willcox, Karen~E}}
  \yr{2004}  \at{Aerodynamic data reconstruction and inverse design using
  proper orthogonal decomposition}.  \jt{AIAA journal}  \bvol{42}~(8),
  \pg{1505--1516}.

\bibitem[Cazemier {\em et~al.\/}(1998)Cazemier, Verstappen \&
  Veldman]{cazemier1998}
{\sc \au{Cazemier, W}, \au{Verstappen, RWCP} \& \au{Veldman, AEP}} \yr{1998}
  \at{Proper orthogonal decomposition and low-dimensional models for driven
  cavity flows}.  \jt{Physics of Fluids (1994-present)}  \bvol{10}~(7),
  \pg{1685--1699}.

\bibitem[Chakravarthy \& Ottino(1996)]{chakravarthy1996mixing}
{\sc \au{Chakravarthy, VS} \& \au{Ottino, JM}} \yr{1996}  \at{Mixing of two
  viscous fluids in a rectangular cavity}.  \jt{Chemical engineering science}
  \bvol{51}~(14),  \pg{3613--3622}.

\bibitem[Chella \& Ottino(1985)]{chella1985fluid}
{\sc \au{Chella, Ravindran} \& \au{Ottino, Julio~M}} \yr{1985}  \at{Fluid
  mechanics of mixing in a single-screw extruder}.  \jt{Industrial \&
  engineering chemistry fundamentals}  \bvol{24}~(2),  \pg{170--180}.

\bibitem[Chella \& Vi{\~n}als(1996)]{chella1996mixing}
{\sc \au{Chella, Ravi} \& \au{Vi{\~n}als, Jorge}} \yr{1996}  \at{Mixing of a
  two-phase fluid by cavity flow}.  \jt{Physical Review E}  \bvol{53}~(4),
  \pg{3832}.

\bibitem[Chien {\em et~al.\/}(1986)Chien, Rising \& Ottino]{chien1986laminar}
{\sc \au{Chien, WL}, \au{Rising, H} \& \au{Ottino, JM}} \yr{1986}  \at{Laminar
  mixing and chaotic mixing in several cavity flows}.  \jt{Journal of Fluid
  Mechanics}  \bvol{170}~(1),  \pg{355--377}.

\bibitem[Coulliette {\em et~al.\/}(2007)Coulliette, Lekien, Paduan, Haller \&
  Marsden]{coulliette2007optimal}
{\sc \au{Coulliette, Chad}, \au{Lekien, Francois}, \au{Paduan, Jeffrey~D},
  \au{Haller, George} \& \au{Marsden, Jerrold~E}} \yr{2007}  \at{Optimal
  pollution mitigation in monterey bay based on coastal radar data and
  nonlinear dynamics}.  \jt{Environmental science \& technology}
  \bvol{41}~(18),  \pg{6562--6572}.

\bibitem[Farazmand \& Haller(2012)]{Farazmand2012}
{\sc \au{Farazmand, Mohammad} \& \au{Haller, George}} \yr{2012}  \at{{Computing
  Lagrangian coherent structures from their variational theory.}}  \jt{Chaos
  (Woodbury, N.Y.)}  \bvol{22}~(1),  \pg{013128}.

\bibitem[Ferrachat \& Ricard(2001)]{ferrachat2001mixing}
{\sc \au{Ferrachat, Sylvaine} \& \au{Ricard, Yanick}} \yr{2001}  \at{Mixing
  properties in the earth's mantle: effects of the viscosity stratification and
  of oceanic crust segregation}.  \jt{Geochemistry, Geophysics, Geosystems}
  \bvol{2}~(4).

\bibitem[Franjione {\em et~al.\/}(1989)Franjione, Leong \&
  Ottino]{franjione1989symmetries}
{\sc \au{Franjione, John~G}, \au{Leong, Chik-Weng} \& \au{Ottino, Julio~M}}
  \yr{1989}  \at{Symmetries within chaos: a route to effective mixing}.
  \jt{Physics of Fluids A: Fluid Dynamics (1989-1993)}  \bvol{1}~(11),
  \pg{1772--1783}.

\bibitem[Froyland {\em et~al.\/}(2010)Froyland, Santitissadeekorn \&
  Monahan]{froyland2010transport}
{\sc \au{Froyland, Gary}, \au{Santitissadeekorn, Naratip} \& \au{Monahan,
  Adam}} \yr{2010}  \at{Transport in time-dependent dynamical systems:
  Finite-time coherent sets}.  \jt{Chaos: An Interdisciplinary Journal of
  Nonlinear Science}  \bvol{20}~(4),  \pg{043116}.

\bibitem[Geyer {\em et~al.\/}(2010)Geyer, Lavery, Scully \&
  Trowbridge]{geyer2010mixing}
{\sc \au{Geyer, W~Rockwell}, \au{Lavery, Andone~C}, \au{Scully, Malcolm~E} \&
  \au{Trowbridge, John~H}} \yr{2010}  \at{Mixing by shear instability at high
  reynolds number}.  \jt{Geophysical Research Letters}  \bvol{37}~(22).

\bibitem[Gharib \& Derango(1989)]{gharib1989liquid}
{\sc \au{Gharib, Morteza} \& \au{Derango, Philip}} \yr{1989}  \at{A liquid film
  (soap film) tunnel to study two-dimensional laminar and turbulent shear
  flows}.  \jt{Physica D: Nonlinear Phenomena}  \bvol{37}~(1-3),
  \pg{406--416}.

\bibitem[Ghia {\em et~al.\/}(1982)Ghia, Ghia \& Shin]{ghia1982high}
{\sc \au{Ghia, UKNG}, \au{Ghia, Kirti~N} \& \au{Shin, CT}} \yr{1982}
  \at{High-re solutions for incompressible flow using the navier-stokes
  equations and a multigrid method}.  \jt{Journal of computational physics}
  \bvol{48}~(3),  \pg{387--411}.

\bibitem[Guckenheimer \& Holmes(1983)]{guckenheimer1983nonlinear}
{\sc \au{Guckenheimer, John} \& \au{Holmes, Philip}} \yr{1983}  \at{Nonlinear
  oscillations, dynamical systems, and bifurcations of vector fields} .

\bibitem[Haller(2015)]{Haller2015}
{\sc \au{Haller, G.}} \yr{2015}  \at{{Langrangian Coherent Structures}}.
  \jt{Annual Review of Fluid Mechanics}  \bvol{47}~(1).

\bibitem[Holmes {\em et~al.\/}(2012)Holmes, Lumley, Berkooz \&
  Rowley]{holmes2012turbulence}
{\sc \au{Holmes, Philip}, \au{Lumley, John~L}, \au{Berkooz, Gal} \& \au{Rowley,
  Clarence}} \yr{2012} {\em Turbulence, coherent structures, dynamical systems
  and symmetry\/}.  \publ{Cambridge university press}.

\bibitem[Hwang {\em et~al.\/}(2005)Hwang, Anderson \& Hulsen]{hwang2005chaotic}
{\sc \au{Hwang, Wook~Ryol}, \au{Anderson, Patrick~D} \& \au{Hulsen, Martien~A}}
  \yr{2005}  \at{Chaotic advection in a cavity flow with rigid particles}.
  \jt{Physics of fluids}  \bvol{17}~(4),  \pg{043602}.

\bibitem[Jana {\em et~al.\/}(1994{\natexlab{{\em a\/}}})Jana, Metcalfe \&
  Ottino]{jana1994experimental}
{\sc \au{Jana, Sadhan~C}, \au{Metcalfe, Guy} \& \au{Ottino, JM}}
  \yr{1994{\natexlab{{\em a\/}}}}  \at{Experimental and computational studies
  of mixing in complex stokes flows: the vortex mixing flow and multicellular
  cavity flows}.  \jt{Journal of Fluid Mechanics}  \bvol{269},  \pg{199--246}.

\bibitem[Jana {\em et~al.\/}(1994{\natexlab{{\em b\/}}})Jana, Tjahjadi \&
  Ottino]{jana1994chaotic}
{\sc \au{Jana, Sadhan~C}, \au{Tjahjadi, Mahari} \& \au{Ottino, JM}}
  \yr{1994{\natexlab{{\em b\/}}}}  \at{Chaotic mixing of viscous fluids by
  periodic changes in geometry: baffled cavity flow}.  \jt{AIChE journal}
  \bvol{40}~(11),  \pg{1769--1781}.

\bibitem[Khakhar {\em et~al.\/}(1987)Khakhar, Franjione \&
  Ottino]{khakhar1987case}
{\sc \au{Khakhar, DV}, \au{Franjione, JG} \& \au{Ottino, JM}} \yr{1987}  \at{A
  case study of chaotic mixing in deterministic flows: the partitioned-pipe
  mixer}.  \jt{Chemical Engineering Science}  \bvol{42}~(12),  \pg{2909--2926}.

\bibitem[{Koopman}(1931)]{Koopman1931}
{\sc \au{{Koopman}, Bernard~O}} \yr{1931}  \at{Hamiltonian systems and
  transformation in hilbert space}.  \jt{Proceedings of the National Academy of
  Sciences}  \bvol{17}~(5),  \pg{315--318}.

\bibitem[Korda {\em et~al.\/}(2018)Korda, Putinar \& Mezi{\'c}]{korda2018data}
{\sc \au{Korda, Milan}, \au{Putinar, Mihai} \& \au{Mezi{\'c}, Igor}} \yr{2018}
  \at{Data-driven spectral analysis of the koopman operator}.  \jt{Applied and
  Computational Harmonic Analysis} .

\bibitem[Koseff \& Street(1984)]{koseff1984}
{\sc \au{Koseff, JR} \& \au{Street, RL}} \yr{1984}  \at{The lid-driven cavity
  flow: a synthesis of qualitative and quantitative observations}.  \jt{Journal
  of Fluids Engineering}  \bvol{106}~(4),  \pg{390--398}.

\bibitem[Large {\em et~al.\/}(1994)Large, McWilliams \&
  Doney]{large1994oceanic}
{\sc \au{Large, William~G}, \au{McWilliams, James~C} \& \au{Doney, Scott~C}}
  \yr{1994}  \at{Oceanic vertical mixing: A review and a model with a nonlocal
  boundary layer parameterization}.  \jt{Reviews of Geophysics}  \bvol{32}~(4),
   \pg{363--403}.

\bibitem[Leong \& Ottino(1989)]{leong1989experiments}
{\sc \au{Leong, CW} \& \au{Ottino, JM}} \yr{1989}  \at{Experiments on mixing
  due to chaotic advection in a cavity}.  \jt{Journal of Fluid Mechanics}
  \bvol{209},  \pg{463--499}.

\bibitem[Ling(1993)]{ling1993effect}
{\sc \au{Ling, FH}} \yr{1993}  \at{The effect of mixing protocol on mixing in
  discontinuous cavity flows}.  \jt{Physics Letters A}  \bvol{177}~(4-5),
  \pg{331--337}.

\bibitem[Ling \& Schmidt(1992)]{ling1992mixing}
{\sc \au{Ling, FH} \& \au{Schmidt, G}} \yr{1992}  \at{Mixing windows in
  discontinuous cavity flows}.  \jt{Physics Letters A}  \bvol{165}~(3),
  \pg{221--230}.

\bibitem[Liu {\em et~al.\/}(1994)Liu, Muzzio \& Peskin]{liu1994quantification}
{\sc \au{Liu, M}, \au{Muzzio, FJ} \& \au{Peskin, RL}} \yr{1994}
  \at{Quantification of mixing in aperiodic chaotic flows}.  \jt{Chaos,
  Solitons \& Fractals}  \bvol{4}~(6),  \pg{869--893}.

\bibitem[Mathew {\em et~al.\/}(2005)Mathew, Mezi{\'c} \&
  Petzold]{mathew2005multiscale}
{\sc \au{Mathew, George}, \au{Mezi{\'c}, Igor} \& \au{Petzold, Linda}}
  \yr{2005}  \at{A multiscale measure for mixing}.  \jt{Physica D: Nonlinear
  Phenomena}  \bvol{211}~(1),  \pg{23--46}.

\bibitem[Meleshko \& Peters(1996)]{meleshko1996periodic}
{\sc \au{Meleshko, VV} \& \au{Peters, GWM}} \yr{1996}  \at{Periodic points for
  two-dimensional stokes flow in a rectangular cavity}.  \jt{Physics Letters A}
   \bvol{216}~(1),  \pg{87--96}.

\bibitem[Mezi{\'c}(2005)]{Mezic2005}
{\sc \au{Mezi{\'c}, Igor}} \yr{2005}  \at{Spectral properties of dynamical
  systems, model reduction and decompositions}.  \jt{Nonlinear Dynamics}
  \bvol{41}~(1-3),  \pg{309--325}.

\bibitem[Mezi\'c(2013)]{Mezic2013analysis}
{\sc \au{Mezi\'c, Igor}} \yr{2013}  \at{Analysis of fluid flows via spectral
  properties of the {Koopman} operator}.  \jt{Annual Review of Fluid Mechanics}
   \bvol{45},  \pg{357--378}.

\bibitem[Mezi{\'c} \& Banaszuk(2004)]{mezic2004comparison}
{\sc \au{Mezi{\'c}, Igor} \& \au{Banaszuk, Andrzej}} \yr{2004}  \at{Comparison
  of systems with complex behavior}.  \jt{Physica D: Nonlinear Phenomena}
  \bvol{197}~(1),  \pg{101--133}.

\bibitem[Mezi{\'c} {\em et~al.\/}(2010)Mezi{\'c}, Loire, Fonoberov \&
  Hogan]{mezic2010new}
{\sc \au{Mezi{\'c}, Igor}, \au{Loire, S}, \au{Fonoberov, Vladimir~A} \&
  \au{Hogan, P}} \yr{2010}  \at{A new mixing diagnostic and gulf oil spill
  movement}.  \jt{Science}  \bvol{330}~(6003),  \pg{486--489}.

\bibitem[Migeon {\em et~al.\/}(2000)Migeon, Texier \&
  Pineau]{migeon2000effects}
{\sc \au{Migeon, C}, \au{Texier, A} \& \au{Pineau, G}} \yr{2000}  \at{Effects
  of lid-driven cavity shape on the flow establishment phase}.  \jt{Journal of
  Fluids and Structures}  \bvol{14}~(4),  \pg{469--488}.

\bibitem[Noack {\em et~al.\/}(2003)Noack, Afanasiev, Morzynski, Tadmor \&
  Thiele]{noack2003hierarchy}
{\sc \au{Noack, Bernd~R}, \au{Afanasiev, Konstantin}, \au{Morzynski, Marek},
  \au{Tadmor, Gilead} \& \au{Thiele, Frank}} \yr{2003}  \at{A hierarchy of
  low-dimensional models for the transient and post-transient cylinder wake}.
  \jt{Journal of Fluid Mechanics}  \bvol{497},  \pg{335--363}.

\bibitem[Ottino {\em et~al.\/}(1992)Ottino, Muzzio, Tjahjadi, Franjione, Jana
  \& Kusch]{ottino1992chaos}
{\sc \au{Ottino, JM}, \au{Muzzio, FJ}, \au{Tjahjadi, M}, \au{Franjione, JG},
  \au{Jana, SC} \& \au{Kusch, HA}} \yr{1992}  \at{Chaos, symmetry, and
  self-similarity- exploiting order and disorder in mixing processes}.
  \jt{Science}  \bvol{257}~(5071),  \pg{754--760}.

\bibitem[Ottino(1989)]{ottino1989kinematics}
{\sc \au{Ottino, Julio~M}} \yr{1989} {\em The kinematics of mixing: stretching,
  chaos, and transport\/}, ,  \vol{vol.~3}.  \publ{Cambridge university press}.

\bibitem[Pai {\em et~al.\/}(2013)Pai, Prakash \& Patnaik]{pai2013numerical}
{\sc \au{Pai, SA}, \au{Prakash, P} \& \au{Patnaik, BSV}} \yr{2013}
  \at{Numerical simulation of chaotic mixing in lid driven cavity: effect of
  passive plug}.  \jt{Engineering Applications of Computational Fluid
  Mechanics}  \bvol{7}~(3),  \pg{406--418}.

\bibitem[Poje {\em et~al.\/}(1999)Poje, Haller \& Mezi{\'c}]{poje1999geometry}
{\sc \au{Poje, Andrew~C}, \au{Haller, George} \& \au{Mezi{\'c}, I}} \yr{1999}
  \at{The geometry and statistics of mixing in aperiodic flows}.  \jt{Physics
  of Fluids (1994-present)}  \bvol{11}~(10),  \pg{2963--2968}.

\bibitem[Prandtl(1904)]{prandtl1904uber}
{\sc \au{Prandtl, Ludwig}} \yr{1904}  \at{Uber flussigkeits bewegung bei sehr
  kleiner reibung}.  \jt{Verhaldlg III Int. Math. Kong}  \pg{pp. 484--491}.

\bibitem[Rom-Kedar {\em et~al.\/}(1990)Rom-Kedar, Leonard \&
  Wiggins]{rom1990analytical}
{\sc \au{Rom-Kedar, V}, \au{Leonard, A} \& \au{Wiggins, S}} \yr{1990}  \at{An
  analytical study of transport, mixing and chaos in an unsteady vortical
  flow}.  \jt{Journal of Fluid Mechanics}  \bvol{214},  \pg{347--394}.

\bibitem[Rowley {\em et~al.\/}(2009)Rowley, Mezi{\'c}, Bagheri, Schlatter \&
  Henningson]{rowley2009}
{\sc \au{Rowley, C.W.}, \au{Mezi{\'c}, I.}, \au{Bagheri, S.}, \au{Schlatter,
  P.} \& \au{Henningson, D.S.}} \yr{2009}  \at{Spectral analysis of nonlinear
  flows}.  \jt{Journal of Fluid Mechanics}  \bvol{641}~(1),  \pg{115--127}.

\bibitem[Schmid(2010)]{Schmid2010}
{\sc \au{Schmid, Peter~J}} \yr{2010}  \at{Dynamic mode decomposition of
  numerical and experimental data}.  \jt{Journal of Fluid Mechanics}
  \bvol{656},  \pg{5--28}.

\bibitem[Shadden \& Taylor(2008)]{shadden2008characterization}
{\sc \au{Shadden, Shawn~C} \& \au{Taylor, Charles~A}} \yr{2008}
  \at{Characterization of coherent structures in the cardiovascular system}.
  \jt{Annals of biomedical engineering}  \bvol{36}~(7),  \pg{1152--1162}.

\bibitem[Sherwood {\em et~al.\/}(2014)Sherwood, Bony \&
  Dufresne]{sherwood2014spread}
{\sc \au{Sherwood, Steven~C}, \au{Bony, Sandrine} \& \au{Dufresne, Jean-Louis}}
  \yr{2014}  \at{Spread in model climate sensitivity traced to atmospheric
  convective mixing}.  \jt{Nature}  \bvol{505}~(7481),  \pg{37--42}.

\bibitem[Solomon \& Mezi{\'c}(2003)]{solomon2003uniform}
{\sc \au{Solomon, TH} \& \au{Mezi{\'c}, Igor}} \yr{2003}  \at{Uniform resonant
  chaotic mixing in fluid flows}.  \jt{Nature}  \bvol{425}~(6956),
  \pg{376--380}.

\bibitem[Stremler \& Chen(2007)]{stremler2007generating}
{\sc \au{Stremler, Mark~A} \& \au{Chen, Jie}} \yr{2007}  \at{Generating
  topological chaos in lid-driven cavity flow}.  \jt{Physics of Fluids
  (1994-present)}  \bvol{19}~(10),  \pg{103602}.

\bibitem[Stroock {\em et~al.\/}(2002)Stroock, Dertinger, Ajdari, Mezi{\'c},
  Stone \& Whitesides]{stroock2002chaotic}
{\sc \au{Stroock, Abraham~D}, \au{Dertinger, Stephan~KW}, \au{Ajdari, Armand},
  \au{Mezi{\'c}, Igor}, \au{Stone, Howard~A} \& \au{Whitesides, George~M}}
  \yr{2002}  \at{Chaotic mixer for microchannels}.  \jt{Science}
  \bvol{295}~(5555),  \pg{647--651}.

\bibitem[Thiffeault(2012)]{thiffeault2012using}
{\sc \au{Thiffeault, Jean-Luc}} \yr{2012}  \at{Using multiscale norms to
  quantify mixing and transport}.  \jt{Nonlinearity}  \bvol{25}~(2),  \pg{R1}.

\bibitem[Tseng \& Ferziger(2001)]{tseng2001mixing}
{\sc \au{Tseng, Yu-heng} \& \au{Ferziger, Joel~H}} \yr{2001}  \at{Mixing and
  available potential energy in stratified flows}.  \jt{Physics of Fluids}
  \bvol{13}~(5),  \pg{1281--1293}.

\bibitem[Valentine {\em et~al.\/}(2012)Valentine, Mezi{\'c},
  Ma{\'c}e{\v{s}}i{\'c}, {\v{C}}rnjari{\'c}-{\v{Z}}ic, Ivi{\'c}, Hogan,
  Fonoberov \& Loire]{valentine2012dynamic}
{\sc \au{Valentine, David~L}, \au{Mezi{\'c}, Igor}, \au{Ma{\'c}e{\v{s}}i{\'c},
  Senka}, \au{{\v{C}}rnjari{\'c}-{\v{Z}}ic, Nelida}, \au{Ivi{\'c}, Stefan},
  \au{Hogan, Patrick~J}, \au{Fonoberov, Vladimir~A} \& \au{Loire, Sophie}}
  \yr{2012}  \at{Dynamic autoinoculation and the microbial ecology of a deep
  water hydrocarbon irruption}.  \jt{Proceedings of the National Academy of
  Sciences}  \bvol{109}~(50),  \pg{20286--20291}.

\bibitem[Vikhansky(2003)]{vikhansky2003chaotic}
{\sc \au{Vikhansky, A}} \yr{2003}  \at{Chaotic advection of finite-size bodies
  in a cavity flow}.  \jt{physics of fluids}  \bvol{15}~(7),  \pg{1830--1836}.

\end{thebibliography}

\end{document}